\documentclass[usenatbib,fleqn]{mnras}
\usepackage{newtxtext,newtxmath}
\usepackage[T1]{fontenc}

\DeclareRobustCommand{\VAN}[3]{#2}
\let\VANthebibliography\thebibliography
\def\thebibliography{\DeclareRobustCommand{\VAN}[3]{##3}\VANthebibliography}

\usepackage{graphicx}	% Including figure files
\usepackage{epstopdf}
\usepackage{amsmath}	% Advanced maths commands
\usepackage{threeparttable}
\usepackage{hyperref}
\usepackage{url}
\defcitealias{vasiliev+2015}{V15}
\makeatletter
\newlength{\abovecaptionskip}%
\makeatother
\graphicspath{{./}}

\newcommand\lsim{\mathrel{\rlap{\lower4pt\hbox{\hskip1pt$\sim$}}
    \raise1pt\hbox{$<$}}}
\newcommand\gsim{\mathrel{\rlap{\lower4pt\hbox{\hskip1pt$\sim$}}
    \raise1pt\hbox{$>$}}}

\title[Mergers and eccentric nuclear disks]{Galactic merger implications for eccentric nuclear disks: a mechanism for disk alignment}

\author[A. Rodriguez et al.]{
Alexander Rodriguez,$^{1}$\thanks{E-mail: alro1526@colorado.edu}
Aleksey Generozov,$^{1}$
Ann-Marie Madigan$^{1}$
\\
$^{1}$JILA and the Astrophysical and Planetary Sciences Department\\
University of Colorado, Boulder\\
}

\date{Accepted XXX. Received YYY; in original form ZZZ}

\pubyear{2020}

\begin{document}
\label{firstpage}
\pagerange{\pageref{firstpage}--\pageref{lastpage}}
\maketitle

\begin{abstract}

The nucleus of our nearest, large galactic neighbor, M31, contains an eccentric nuclear disk--a disk of stars on eccentric, apsidally-aligned orbits around a supermassive black hole (SMBH).
Previous studies of eccentric nuclear disks considered only an isolated disk, and did not study their dynamics under galaxy mergers (particularly a perturbing SMBH). Here, we present the first study of how eccentric disks are affected by a galactic merger.  We perform $N$-body simulations to study the disk under a range of different possible SMBH initial conditions. A second SMBH in the disk always disrupts it, but more distant SMBHs can shut off differential precession and stabilize the disk. This results in a more aligned disk, nearly uniform eccentricity profile, and suppression of tidal disruption events compared to the isolated disk. We also discuss implications of our work for the presence of a secondary SMBH in M31.\\
\end{abstract}


\section{Introduction} \label{sec:intro}
Certain galactic nuclei contain eccentric nuclear disks: disks of stars on apsidally-aligned, eccentric orbits around a  supermassive black hole (SMBH). In fact, our nearest large galactic neighbor, M31, contains a double nucleus with two brightness peaks offset from the SMBH \citep{lauer+1993}, which are best explained by a single eccentric nuclear disk \citep{tremaine1995}. \citet{Gruzinov2020} show that this disk may be an example of a
lopsided thermodynamic equilibrium of a rotating star cluster around a SMBH.
Additionally many nearby galaxies contain double nuclei, offset nuclei, or nuclei with a central minimum \citep{lauer+2005}, which could be explained by such disks \citep{madigan+2018}.

Eccentric disks are kept stable via secular (orbit-averaged) torques \citep{madigan+2018}. Orbits that are pushed ahead of the disk are torqued to higher eccentricities. This slows their precession rate so that the bulk of the disk can catch up with these wayward orbits. However, stars may also be torqued to nearly radial, loss-cone orbits that pierce the tidal radius ($r_t$), and experience a tidal disruption event (TDE; \citealt{rees1988}). Eccentric disks can produce TDEs at a rate that is two or more orders magnitude higher than two-body relaxation in spherical nuclei (\citealt{madigan+2018}; see \citealt{stone+2020} for a review of the classical loss cone theory).

Previous studies of eccentric disks have only considered isolated disks around a single SMBH. In fact a second SMBH may also be present, considering that eccentric disks can form in gas rich galaxy mergers \citep{hopkins&quataert2010}.

Here, we use $N$-body simulations to investigate disk dynamics in the presence of a perturbing SMBH. We explore different perturber inclinations, semi-major axes, and masses. Since a higher mass SMBH with a small semi-major axis would have a larger gravitational influence, these conditions would produce the most significant changes to the disk dynamics. At larger semi-major axes and smaller masses the disk would more closely resemble the isolated case. 

The paper is organized as follows: In \S~\ref{sec:Modeling Disk SMBH Interactions} we summarize our simulation setup. In \S~\ref{sec:Results} we show the results of our simulations. We find that certain perturber conditions shut off differential precession in the disk, increasing its alignment.  Furthermore, we investigate implications for TDEs.
In \S~\ref{sec:Discussion} we present broader astrophysical implications of our findings. In \S~\ref{sec:Summary}, we summarize our results.

\section{Modeling Disk-SMBH Interactions}
\label{sec:Modeling Disk SMBH Interactions}

\subsection{Simulations}
\label{subsec:Simulations}
We model eccentric disks with $N$-body simulations. In particular, we simulate disks with 100 equal mass particles, orbiting a central SMBH with 200 times the disk mass. The stars initially have an eccentricity of 0.7. The orbits all have similar spatial orientations, with a small scatter ($\sim 2^{\circ}$) in inclination, argument of periapsis, and the longitude of the ascending node.\footnote{Scatter in these angles is introduced using the procedure described in \citet{foote+2020}.} The outer semi-major axis of the disk is twice the inner semi-major axis, and each star's semi-major axis is drawn from a log-uniform distribution. We measure lengths and times in units of the semi-major axis and orbital period at the inner disk edge respectively. All masses are normalized to the central mass. We integrate the disk for 500 orbits using the IAS15 integrator \citep{rein.spiegel2015} within the REBOUND integration framework \citep{rein.liu2012}. (Though we do perform a handful of longer integrations up to 5000 orbits.) For better statistics we run four simulations for each set of parameters.

Next we introduce a perturbing SMBH to our system. We fix the perturber on a circular orbit and explore how the evolution of the disk varies with the perturber mass, semi-major axis, and inclination. We consider two grids of perturber parameters summarized in Tables~\ref{tab:params1} and~\ref{tab:params2}. The first grid consists of relatively small close--in SMBHs. The second includes more distant and massive perturbers, with a fixed inclination of $180^{\circ}$. The motivation for this choice of parameter space is explained in \S~\ref{subsec:Simulations with massive,distant perturbers}. 

Previous work has found the eccentric disks efficiently produce stellar tidal disruption events (TDEs) \citep{madigan+2018, wernke&madigan2019, foote+2020}.
SMBHs can disrupt stars that pass within the tidal radius, viz.
\begin{equation}
\label{r_t}
    r_{t}=\left(\frac{M_{0}}{M_{*}}\right)^\frac{1}{3}R_{*}
\end{equation}
where $M_{0}$, $M_{*}$, and $R_{*}$ are the black hole mass, stellar mass, and stellar radius respectively \citep{rees1988}. To set a tidal radius in our simulation we have to choose an overall length scale. Here, we assume the inner disk edge is at 0.05 pc. Then for a sun-like star and a $10^7 M_{\odot}$ SMBH the tidal radius would be $\sim 10^{-4}$ times the inner disk edge. We record a TDE whenever a star passes within this distance of the central SMBH, using REBOUND's built--in collision detection capabilities.

\begin{table}
\caption{\label{tab:params1} Perturber Parameters for our First Simulation Grid.}
\begin{threeparttable}
\begin{tabular*}{\columnwidth}{@{\extracolsep{\fill}}lccc}
    \hline
    \textbf{Parameter} & \textbf{Lower Bound} & \textbf{Upper Bounder} & $\mathbf{n_{points}}$ \\ 
    \hline
     $a_p$  & 2 & 6 & 5\\ \hline
 $\log(M_{p})$ & $\log(5\times 10^{-4})$  &$\log(5\times 10^{-2})$ & 5 \\ \hline
 $i_{p}$ & 0$^{\circ}$ & 180$^{\circ}$ & 5\\ 
      \hline
\end{tabular*}
\begin{tablenotes}
\setlength\labelsep{0pt}
\item Perturber semi-major axes ($a_p$), log masses ($\log(M_p)$), and inclinations ($i_p$) for our first grid of SMBH--disk simulation. For each parameter we use a few evenly spaced grid points between the lower and upper bounds in the last two columns. The number of grid points is indicated by the  ``$n_{\rm points}$'' column.
\end{tablenotes}
\end{threeparttable}
\end{table}

\begin{table}
\caption{\label{tab:params2} Perturber Parameters for our Second Simulation Grid.}
\begin{threeparttable}
\begin{tabular*}{\columnwidth}{@{\extracolsep{\fill}}lccc}
    \hline
    \textbf{Parameter} & \textbf{Lower Bound} & \textbf{Upper Bounder} & $\mathbf{n_{points}}$ \\ 
    \hline
     $a_p$  & 6 & 12 & 7\\ \hline
    $\log(M_{p})$  & $\log(5\times 10^{-2})$  & $\log(5\times 10^{-1})$  & 10
    \\
    \hline
\end{tabular*}
\begin{tablenotes}
\item Parameters for our second grid of SMBH--disk $N$-body simulations. The format is the same as Table~\ref{tab:params1}, except in this case the perturber inclination is fixed at 180$^{\circ}$.
\end{tablenotes}
\end{threeparttable}
\end{table}

\subsection{Quantifying disk alignment}
\label{subec:Quantifying Disk Alignment}

Following \citet{madigan+2018} the apsidal-alignment of a disk can be quantified using the spread in the disk orbits' eccentricity vectors, viz.
\begin{equation}
    {\vec{e}} = \frac{\vec{v}\times\vec{j}}{G M\textsubscript{0}} - \hat{r}, 
\end{equation}
where ${\vec{j}}={\vec{r}}\times\vec{v}$ is the specific angular momentum. This vector points from the apocenter to the pericenter of the orbit, and its magnitude is the eccentricity. For a disk in the $xy$ plane, 
orbital alignment can be measured via the spread in $i_e\equiv \arctan(\frac{e_y}{e_x})$. Here  $e_x$ and $e_y$ are the $x$ and $y$ components of the eccentricity vector. We use the circular standard deviation of $i_e$ ($\sigma_{i_e}$) to quantify the angular spread of the disk.

Orbital alignment can also be evaluated via the Rayleigh dipole statistic
\citep{rayleigh1919}, viz.
\begin{equation}
\label{eq:Sn}
S_{\rm n} = \frac{\left|\sum_{i=1}^{n_{\rm bound}} \hat{e}_{i}\right|}{n_{\rm bound}},
\end{equation}
$n_{\rm bound}$ is the number of bound stars. $S_{\rm n}$ measures the clustering of eccentricity vectors in three dimensional space. Analogously,

\begin{equation}
\label{eq:Snxy}
    S_{\rm n, xy}= \frac{\left|\sum_{i=1}^{n_{\rm bound}} <\hat{e}_{i,x},\hat{e}_{i,y},0>\right|}{n_{\rm bound}}
\end{equation}
measures clustering of the eccentricity vectors in the xy plane, and
\begin{equation}
\label{eq:Snz}
    S_{n,z}= \frac{\left| \sum_{i}^{n_{\rm bound}} \hat{e}_{i, z}\right|}{n_{\rm bound}}
\end{equation}
measures clustering along the z--axis (i.e. along the angular momentum of the disk). 
For isotropically distributed eccentricity vectors, these statistics will be close to 0. Conversely, a dipole statistic of 1 indicates perfect alignment.  For the initial conditions in $\S$~\ref{sec:Modeling Disk SMBH Interactions}, $S_{\rm n, xy}\approx 1$ and $S_{n,z}\approx 0$. 

We use both $\sigma_{i_{e}}$ and $S_{\rm n, xy}$ to make sure our results are statistic-independent. Additionally, $S_{n}$ can be used to measure clustering along arbitrary axes while $\sigma_{i_{e}}$ only measures clustering in the plane of the disk.

\section{Results}
\label{sec:Results}

\subsection{Simulations with close perturbers}
\label{subsec:Simulations with close Perturbers}

\begin{figure}
\includegraphics[width=.45\textwidth]{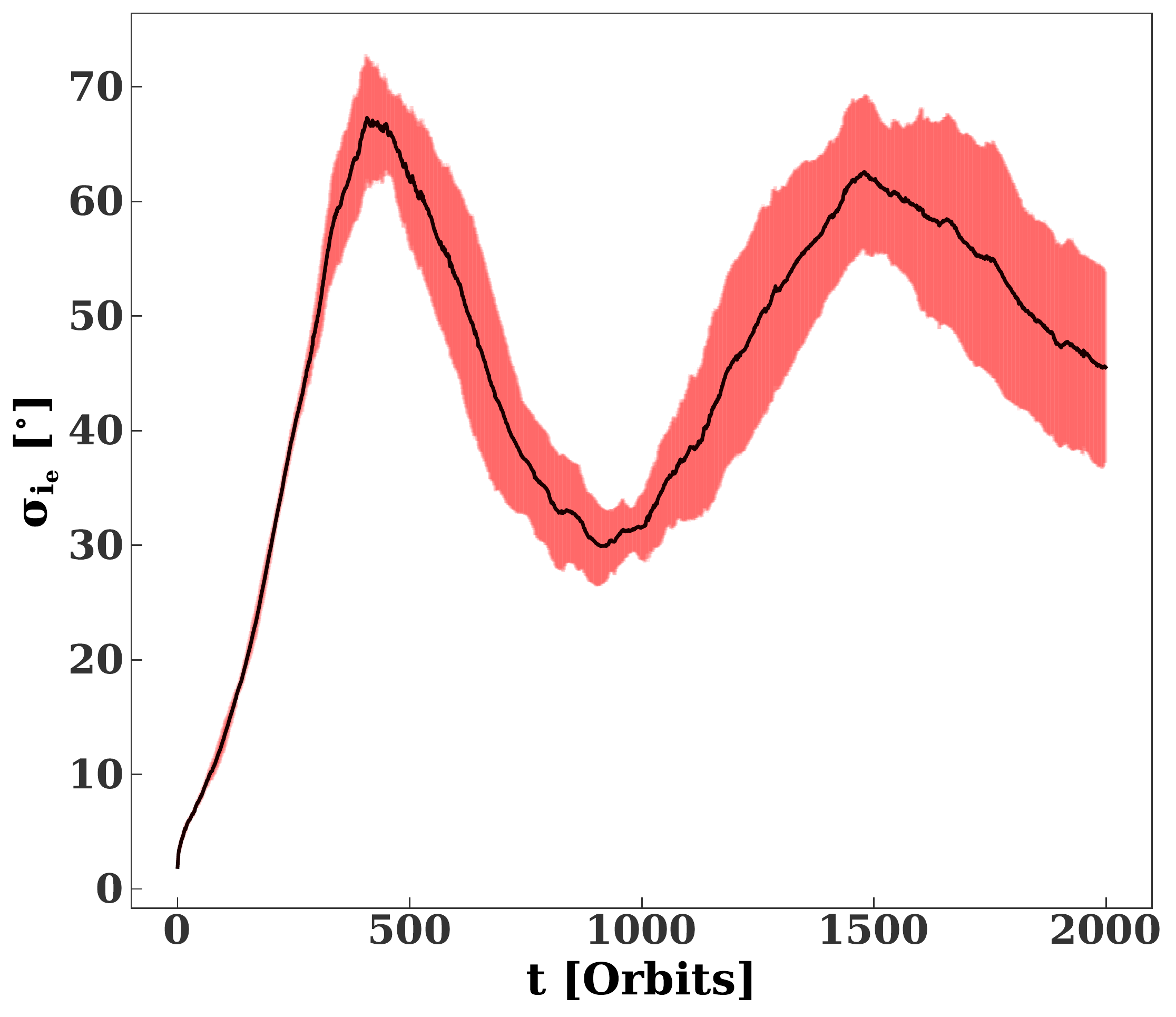}
\caption{\label{fig:2000orbits} The circular standard deviation of $i_e$, $\sigma_{i_{e}}$ (see \S~\ref{subec:Quantifying Disk Alignment}), as a function of time for an isolated disk. The standard deviation oscillates, indicating periodic variations in the strength of disk alignment.
The line is an average over four different simulations. The red area region is the standard deviation over these simulations.}
\end{figure}

\begin{figure*}
\includegraphics[width=.45\textwidth]{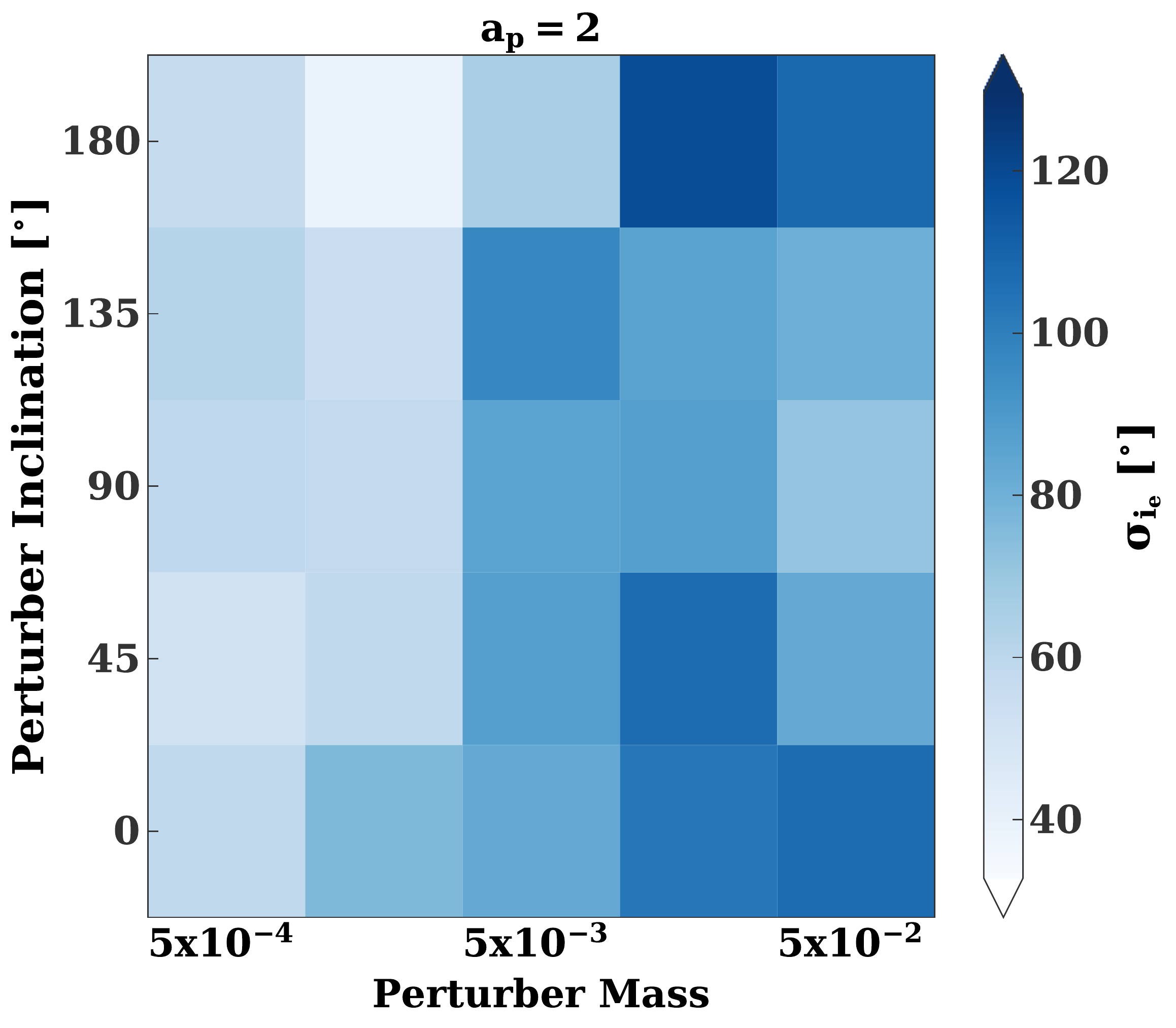}
\includegraphics[width=.45\textwidth]{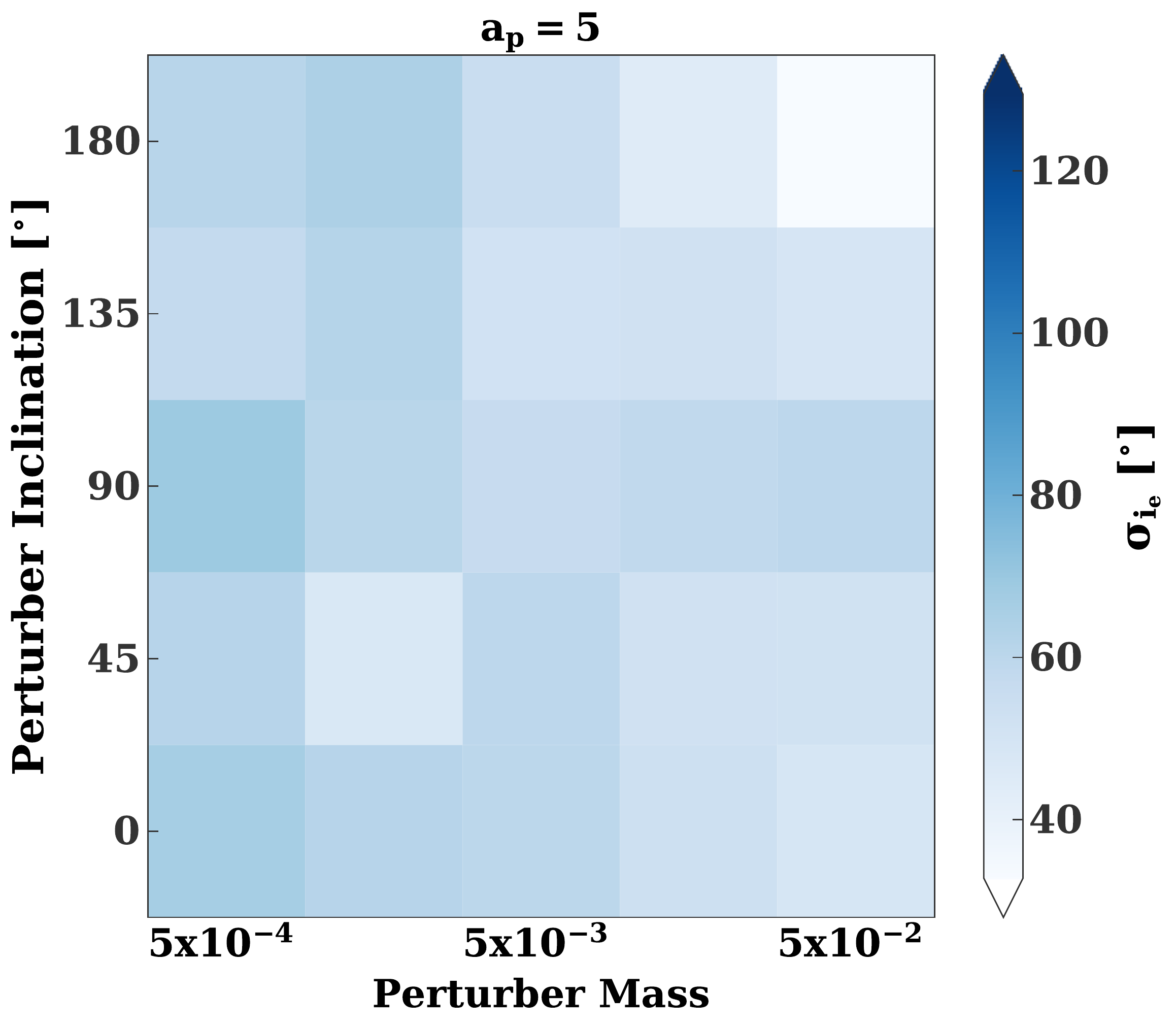}
\includegraphics[width=.45\textwidth]{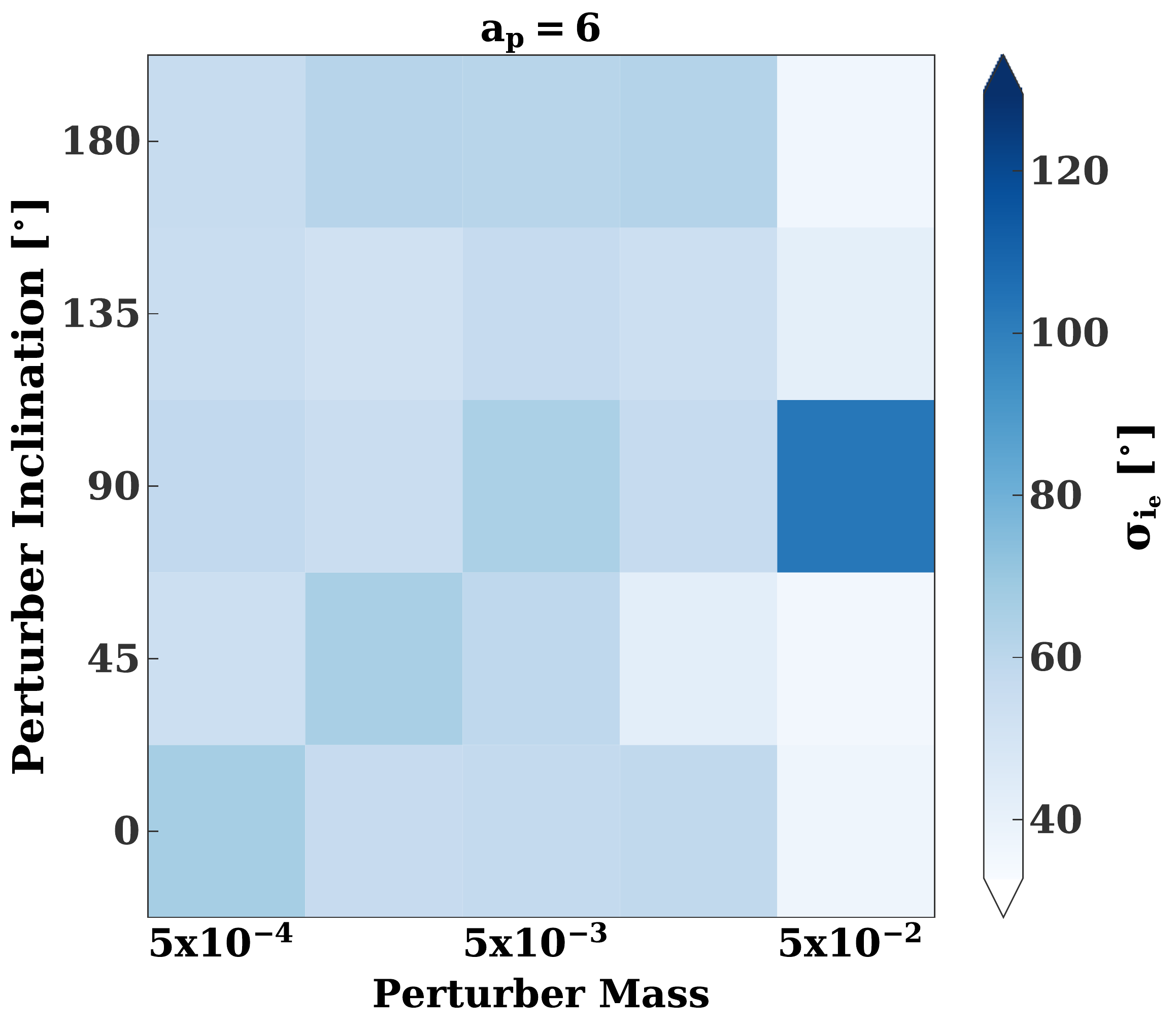}
\caption{\label{fig:ie_smas} Top left panel: Circular standard deviation of $i_e$ after 500 orbital periods as a function of perturber mass and inclination. Here, the perturber orbits inside of the disk with a semi-major axis, $a_p$, of 2. A larger standard deviation indicates a less aligned, more axisymmetric disk. Increasing perturber mass generally increases the standard deviation, and coplanar orbits lead to the most axisymmetric disks as shown by the top and bottom rows. Top right panel: Same as the top left panel, except the perturber semi-major axis is 5. Here, the perturber has a much weaker effect on disk alignment, so that the spread is similar to an isolated disk except when under the influence of massive, retrograde, coplanar perturbers. Bottom panel: Same as top panels, except the perturber semi-major axis is now 6. Remarkably, increasing the perturber mass in this case increases the alignment of the disk for all inclinations except 90$^{\circ}$.}
\end{figure*}

In isolated disk simulations, the disk alignment ($\sigma_{i_{e}}$) oscillates with a period of $\sim$ 1000 orbits; see Figure~\ref{fig:2000orbits}. The oscillations are due to the retrograde precession of the outer disk. When the outer part has precessed far from the inner part, the angular spread will be large and when it precesses back in line with the disk the angular spread will be small again. 
As shown below, disk alignment can be suppressed or amplified by a perturbing SMBH. 

Figure~\ref{fig:ie_smas} illustrates how the small close-in perturbers in Table~\ref{tab:params1} affect disk alignment.
This figure shows the spread of $i_e$ after 500 orbital periods for different perturber semi-major axes, inclinations, and masses. For comparison, $\sigma_{i_e}$ is 65$^{\circ} \pm 10^{\circ}$ in our isolated disk simulations after 500 orbits. The uncertainty is the standard deviation of $\sigma_{i_e}$ measurements from different simulations.

 The top left panel of Figure~\ref{fig:ie_smas} shows that perturbers inside the disk increase the spread of $i_e$ above that of an isolated disk.  Note that these figures only show the alignment of bound stars, and in this case $64 \pm 3$ stars are unbound (out of a total of 100). Inside the disk, the perturber simply ejects the majority of stars and scatters the eccentricity vectors of the remaining disk stars. 
 
 The top right panel of Figure~\ref{fig:ie_smas} shows how removing the perturber from the disk (to a semi-major axis of 5) dramatically reduces the scatter of its eccentricity vectors. Massive perturbers on retrograde orbits can in fact \textit{stabilize} the disk, decreasing the standard deviation of $i_e$ below that of an isolated disk. In all cases at a semi-major axis of 5, the spread of $i_e$ is less than that at closer semi-major axes, indicating a reduction in the effects of the secondary. 
However the retrograde massive perturber is the only one with a statistically significant difference from the isolated case, being $\sim$ 3 times more aligned. Furthermore, the number of  unbound stars at this semi-major axis is only $7 \pm 3$.
 
\begin{figure}
{
\includegraphics[width=.45\textwidth]{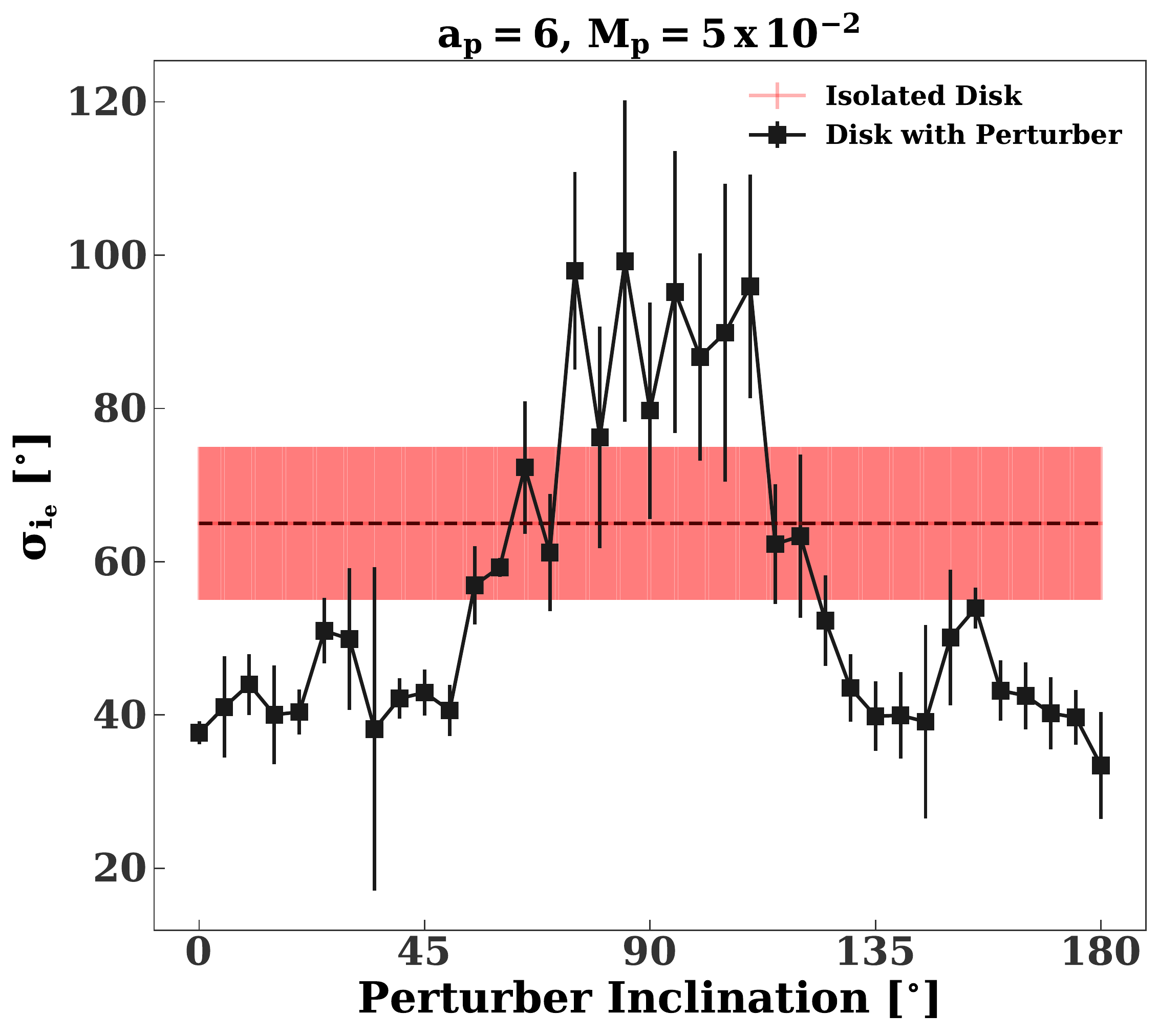}
\caption{\label{fig:manyincs}
The black line shows the circular standard deviation of $i_e$ after 500 orbital periods, as a function of perturber inclination. The perturber mass and semi-major axis are $M_p = 5\times 10^{-2}$ and $a_p = 6$ respectively. The red line and the shaded area are the average and standard deviation of four different isolated disk simulations after 500 orbits as described in~\ref{subsec:Simulations}. As expected from the bottom panel of Figure~\ref{fig:ie_smas}, the standard deviation has a maximum at $i_p = 90^{\circ}$.
}
}
\end{figure}

In the bottom panel of Figure~\ref{fig:ie_smas} the perturber semi-major axis is 6, and the number of unbound stars has dropped further to $3 \pm 1$. Both prograde and retrograde coplanar perturbers increase the alignment of the disk.
Meanwhile, high-mass, orthogonal perturbers ($i_p=90^{\circ}$) cause the disk to become less aligned. Figure~\ref{fig:manyincs} shows the final spread in $i_e$ across a finer grid of inclinations for $a_p=6$, $M_p=0.05$ (the most massive perturber from the bottom panel of  Figure~\ref{fig:ie_smas}). 
As in the bottom panel of Figure \ref{fig:ie_smas}, the most prominent alignment occurs for nearly co-planar perturbers. Reducing the projection of the perturber orbit in the disk plane decreases the disk alignment.

\begin{figure}
{
\includegraphics[width=.45\textwidth]{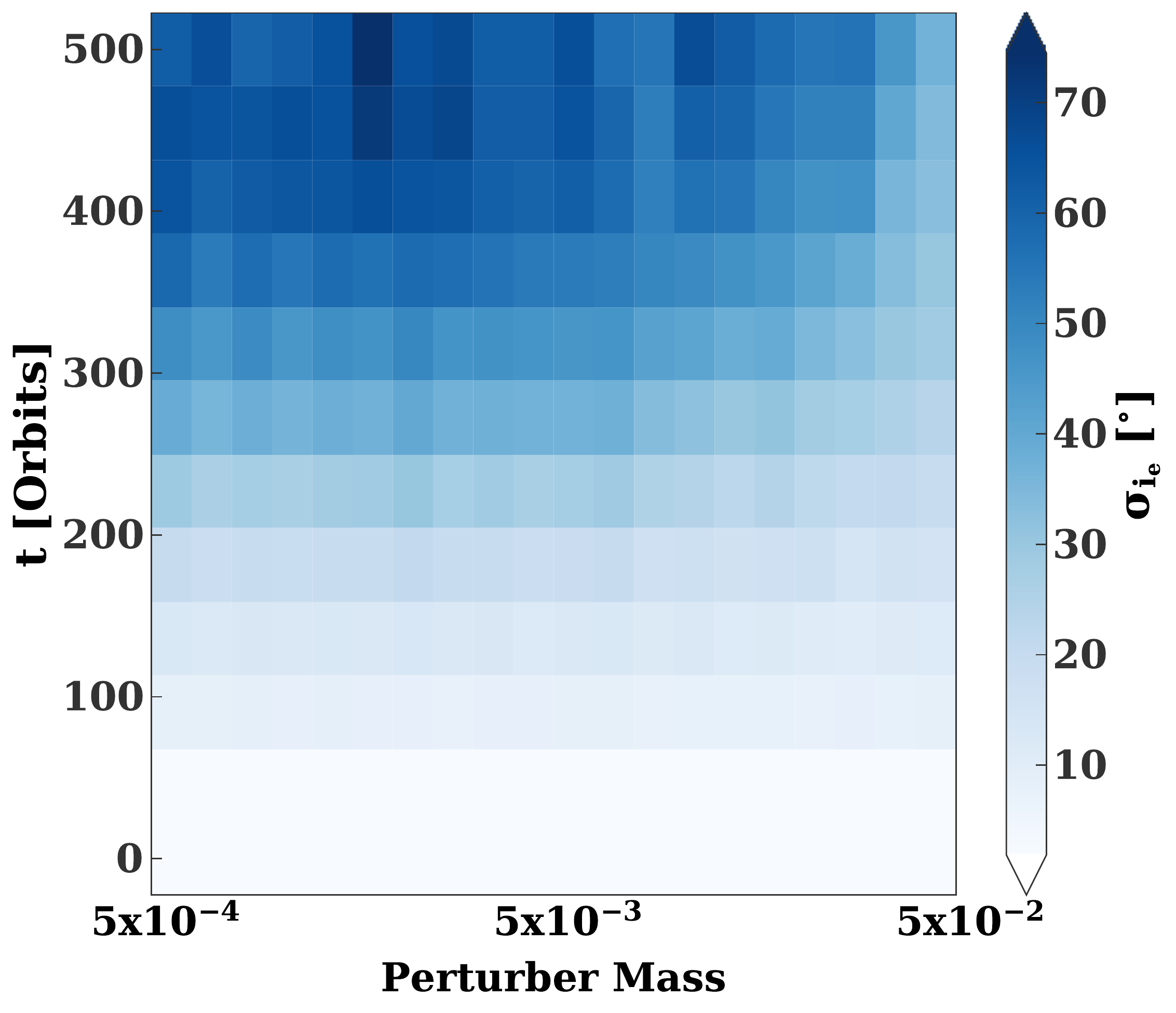}
\caption{\label{fig:ieTime} Circular standard deviation of $i_e$, as a function of time and perturber mass for $a_{p}=6$ and $i_{p}=180^{\circ}$. 
The disk alignment begins evolving after half a secular time. Disks with massive perturbers retain their alignment throughout the simulation. 
}
}
\end{figure}

Figure~\ref{fig:ieTime} shows how disk alignment evolves with time for different mass perturbers (with $i_p=180^{\circ}$, $a_p=6$).
In all cases, the disk starts with near perfect alignment. At the last (500th) orbit, the simulations with the highest mass perturber retain the strongest alignment.  Differences in alignment between perturbers appear after after half a secular time,
\begin{equation}
    t_{\rm sec}=\frac{M_{0}}{M_{\rm disk}} P_{\rm disk},
    \label{eq:tsec}
\end{equation}
where $P_{\textrm{disk}}$ is the orbital period at the inner edge of the disk and $M_{\rm disk}$ is the total disk mass. For the disks simulated here, $t_{\rm sec}$=200 $P_{\rm disk}$.
By 200 orbits, $\sigma_{i_{e}}$ differs by $5^{\circ}$ between $M_p=5\times10^{-4}$ and $5\times10^{-2}$. 

Overall, at large semi-major axes, nearly co-planar perturbers result in more aligned disks. To explore this effect further, we extend our study to larger perturber masses and semi-major axes, focusing on coplanar perturbers.

\subsection{Simulations with massive, distant perturbers}
\label{subsec:Simulations with massive,distant perturbers}

\subsubsection{Analysis using $\sigma_{i_e}$}
\label{subsubsec:Analysis using sigma_ie}

The most distant and massive perturbers in Table~\ref{tab:params1} have the largest effect on the angular extent of the disk. Thus, we increase the perturber semi-major axis and mass to find the limits of this effect. We focus on coplanar perturbers as these had  the most significant effect on disk alignment in $\S$~\ref{subsec:Simulations with close Perturbers}. Specifically we fix the perturber inclination to $180^{\circ}$ though we return to prograde perturbers in \S~\ref{subsec:Prograde vs. retrograde orbits}. The perturber parameters in this section are summarized in Table~\ref{tab:params2}.

\begin{figure}
{
\includegraphics[width=.45\textwidth]{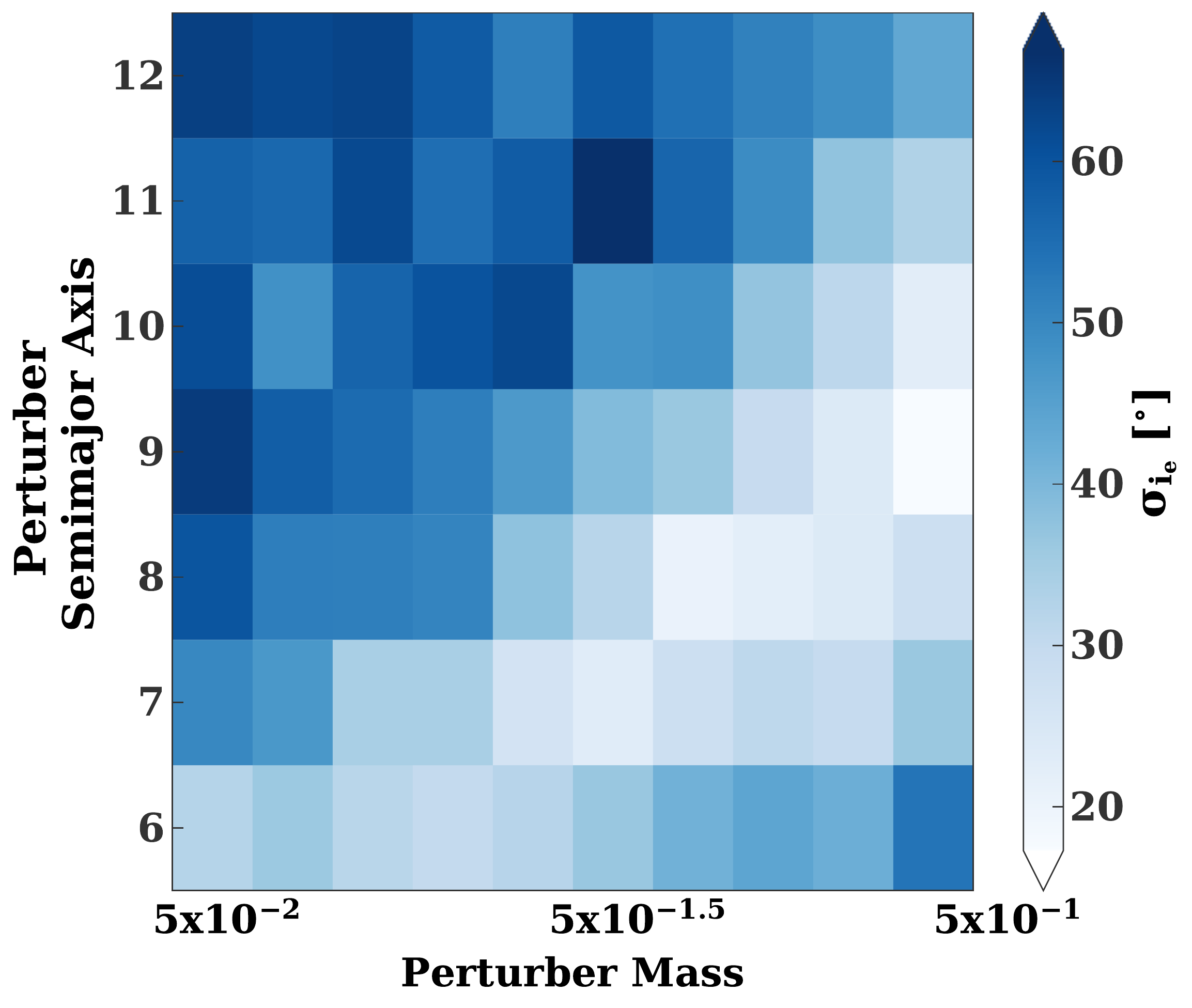}
\caption{\label{fig:ie4} Circular standard deviation of $i_e$ as a function of perturber mass and semi-major axis for the more distant, retrograde perturber in Table~\ref{tab:params2}. The highest mass perturbers destabilize the disk at smaller semi-major axes but the stabilizing effect from $\S$~\ref{subsec:Simulations with close Perturbers} is present at larger semi-major axes.  In fact, for $M_p=0.5$, the disk alignment peaks at a semi-major axis of 9, where the standard deviation of $i_e$ is just $17^{\circ}$.
}
\par
}
\end{figure}

Figure~\ref{fig:ie4} summarizes the results of these simulations. For the largest semi-major axes, the perturber has minimal effect on disk alignment and the standard deviation of $i_e$ approaches that of an isolated disk. Disk alignment does not vary monotonically with perturber separation, and is in general minimized at intermediate semimajor axes.
Within this grid, the standard deviation of $i_e$ is minimized at $M_p=0.5, a_p=9$, where disks are $\sim$ 4 times more aligned than the isolated case.
 We abbreviate these perturber parameters as ``$p_{\rm ideal}$.''

\begin{figure}
{
\includegraphics[width=.45\textwidth]{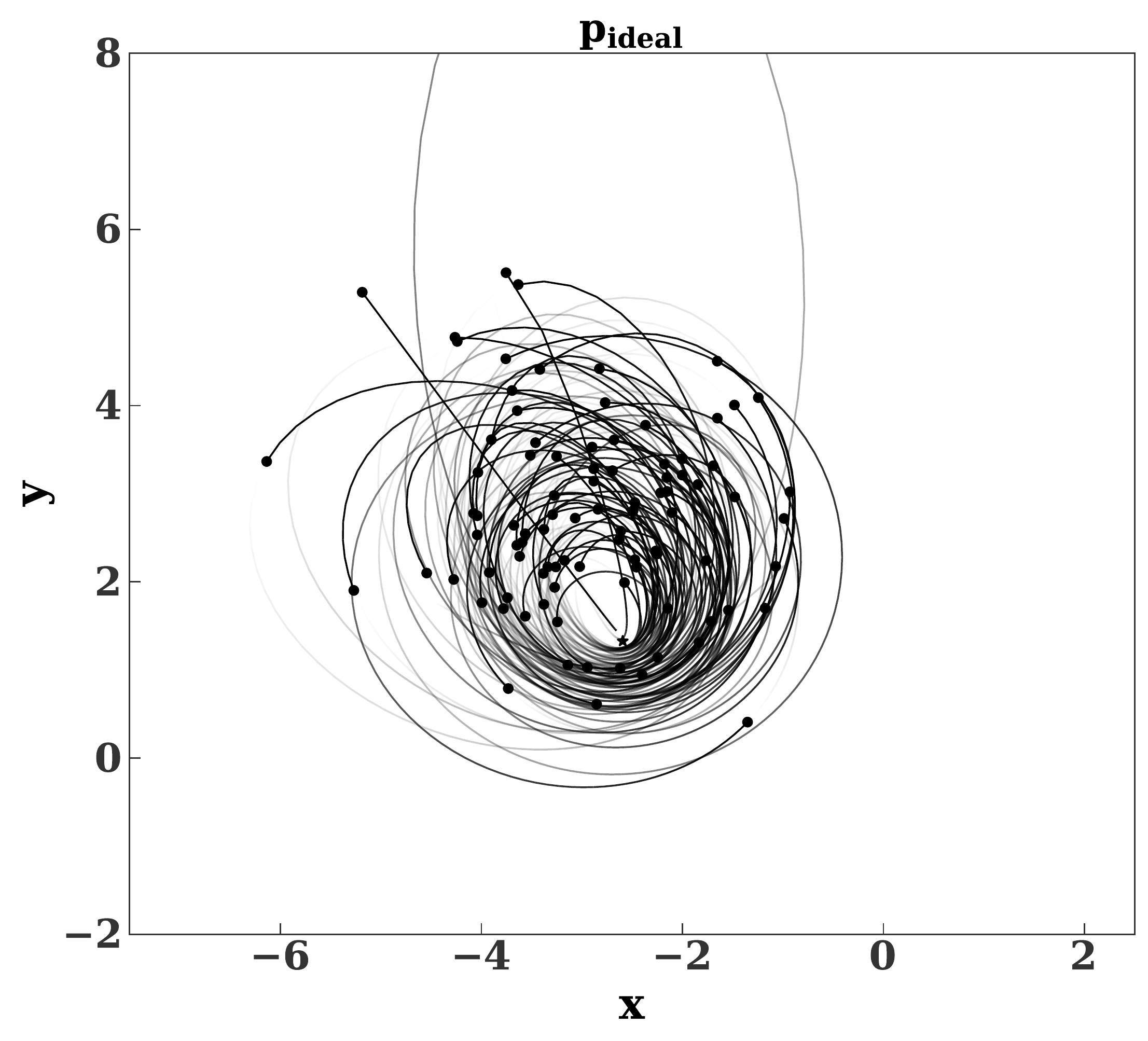}
\includegraphics[width=.45\textwidth]{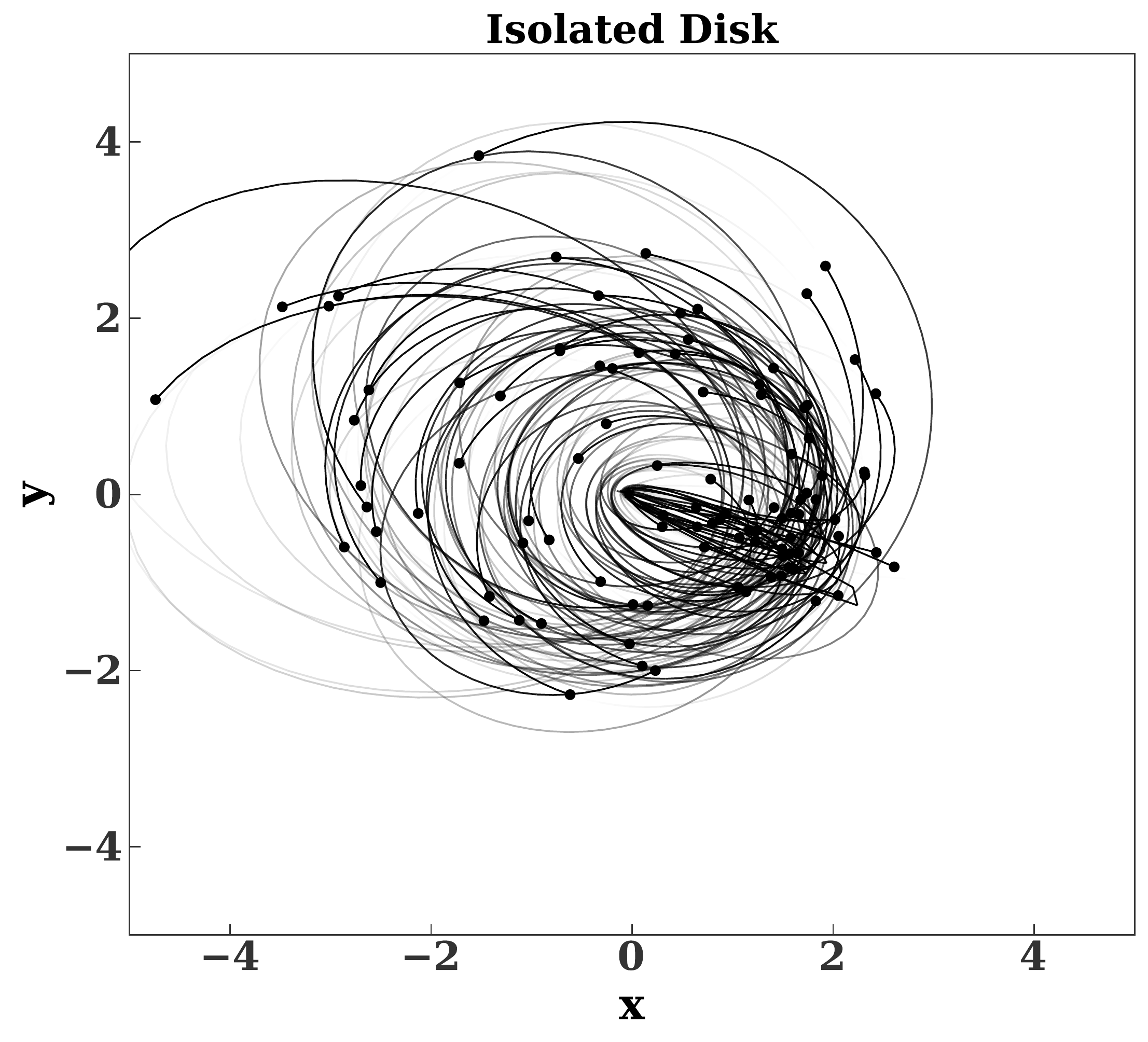}
\caption{\label{fig:orbitplots}  Top panel: snapshot of a $p_{\rm ideal}$ simulation ($M_p=0.5, a_p=9, i_{p}=180^{\circ}$), after 500 orbits at the inner disk edge. Bottom panel: snapshot from an isolated disk simulation after 500 orbits at the inner disk edge. The disk is more aligned and eccentric in the top panel. Note the axes are shifted, but their aspect ratio is the same in both panels.}}
\end{figure}

 The top and bottom panels of Figure~\ref{fig:orbitplots} show the final snapshots of disk orbits from an isolated disk and a $p_{\rm ideal}$ simulation respectively, confirming that the latter is more aligned. 

\begin{figure}
{
\includegraphics[width=.45\textwidth]{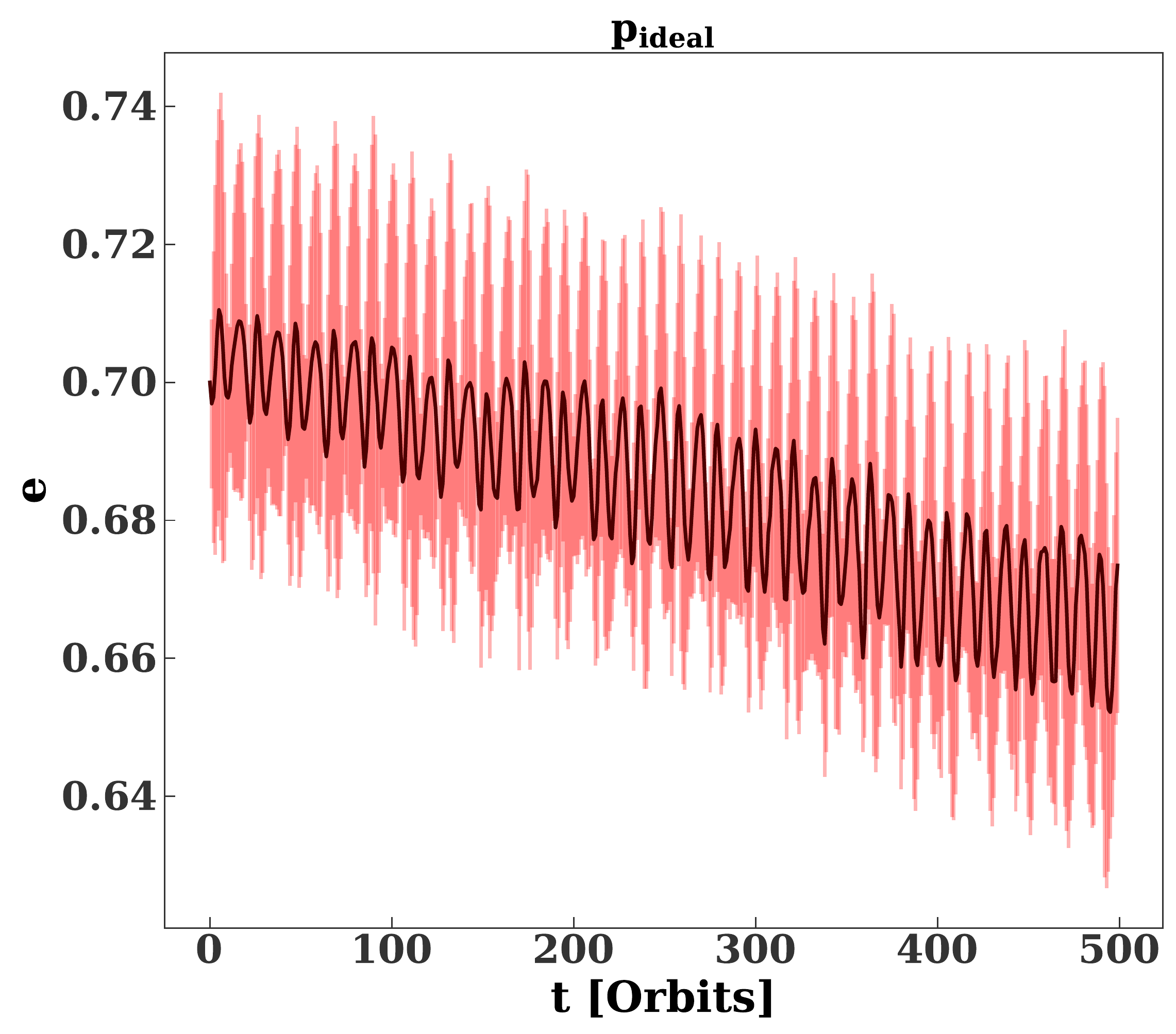}
\includegraphics[width=.45\textwidth]{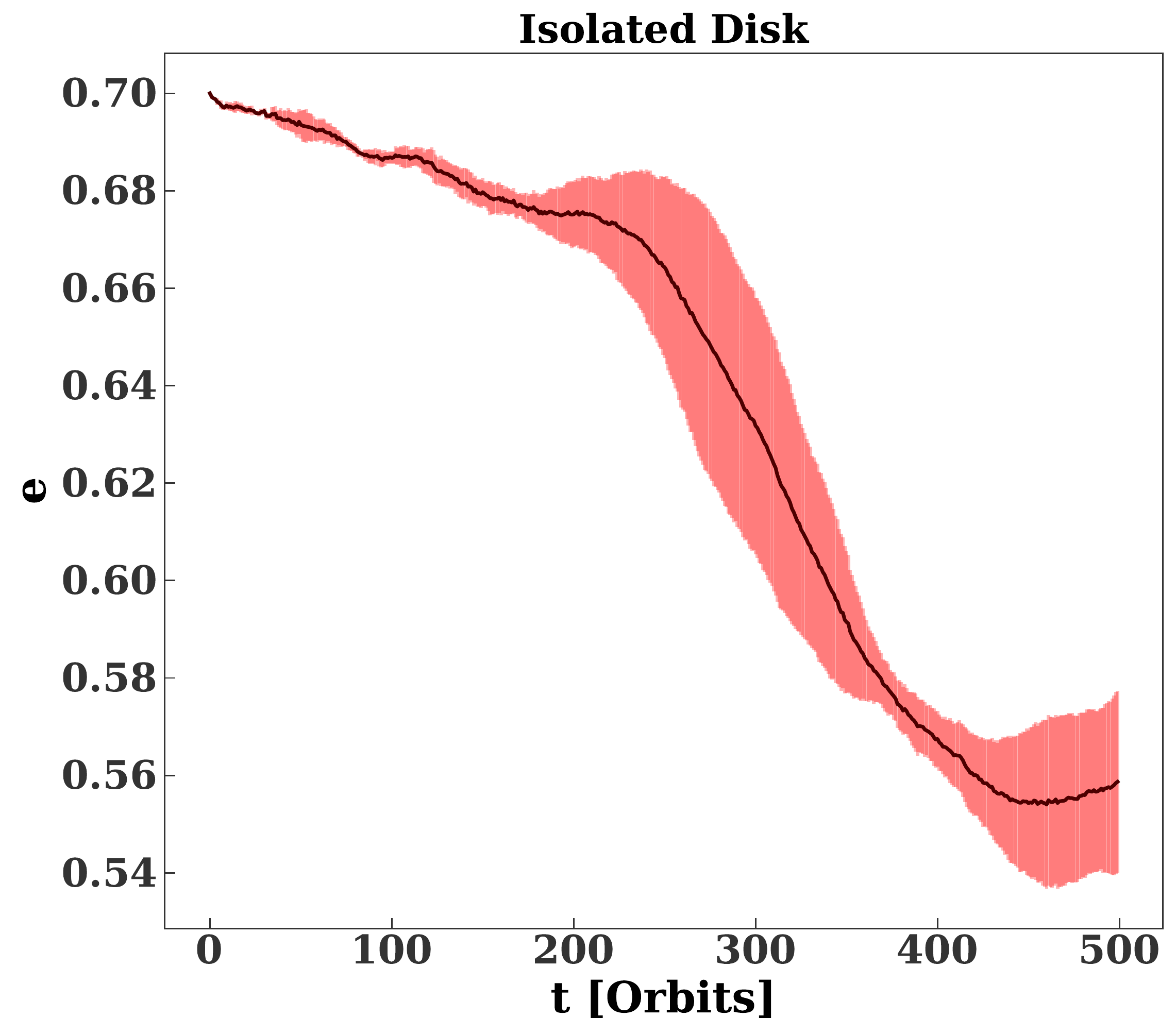}
\caption{\label{fig:eccsplots} Top panel: average eccentricity as a function of time for a disk interacting with the $p_{\rm ideal}$ perturber ($M_p=0.5, a_p=9, i_{p}=180^{\circ}$). The line is an average over the four different simulations as explained in~\ref{subsec:Simulations} (i.e. we compute the mean eccentricity of all bound stars as a function of time in four different simulations and average the results.) The shaded area is the standard deviation. Bottom panel: same for an isolated disk. The disk eccentricity slowly declines until one secular time ($\sim 200$  orbits; equation~\ref{eq:tsec}), after which the rate of decline increases before leveling off at $\sim$ 500 orbits (with a point of inflection at $\sim$ 370 orbits).
}}
\end{figure}

The $p_{\rm ideal}$ disk  has a larger mean eccentricity than the isolated case. 
The top and bottom panels of Figure~\ref{fig:eccsplots} show the mean eccentricity of the disk as a function of time for these simulations.  The isolated disk's eccentricity slowly declines until one secular time (equation~\ref{eq:tsec}), after which the rate of decline increases before leveling off at $\sim$ 500 orbits, with a point of inflection at $\sim$ 370 orbits. For $p_{\rm ideal}$, the eccentricity evolves similarly at early times. However, the slope doesn't change at 200 orbits. Rather, the decline is constant (modulo oscillations due to perturber-induced jitter in the center of mass of the disk-SMBH system), resulting in a larger final (mean) eccentricity for $p_{\rm ideal}$. Moreover, the final mean disk eccentricity for $p_{\rm ideal}$ is higher (and closer to initial conditions) than in any other perturbed disk simulation.

\begin{figure}
{
\includegraphics[width=.45\textwidth]{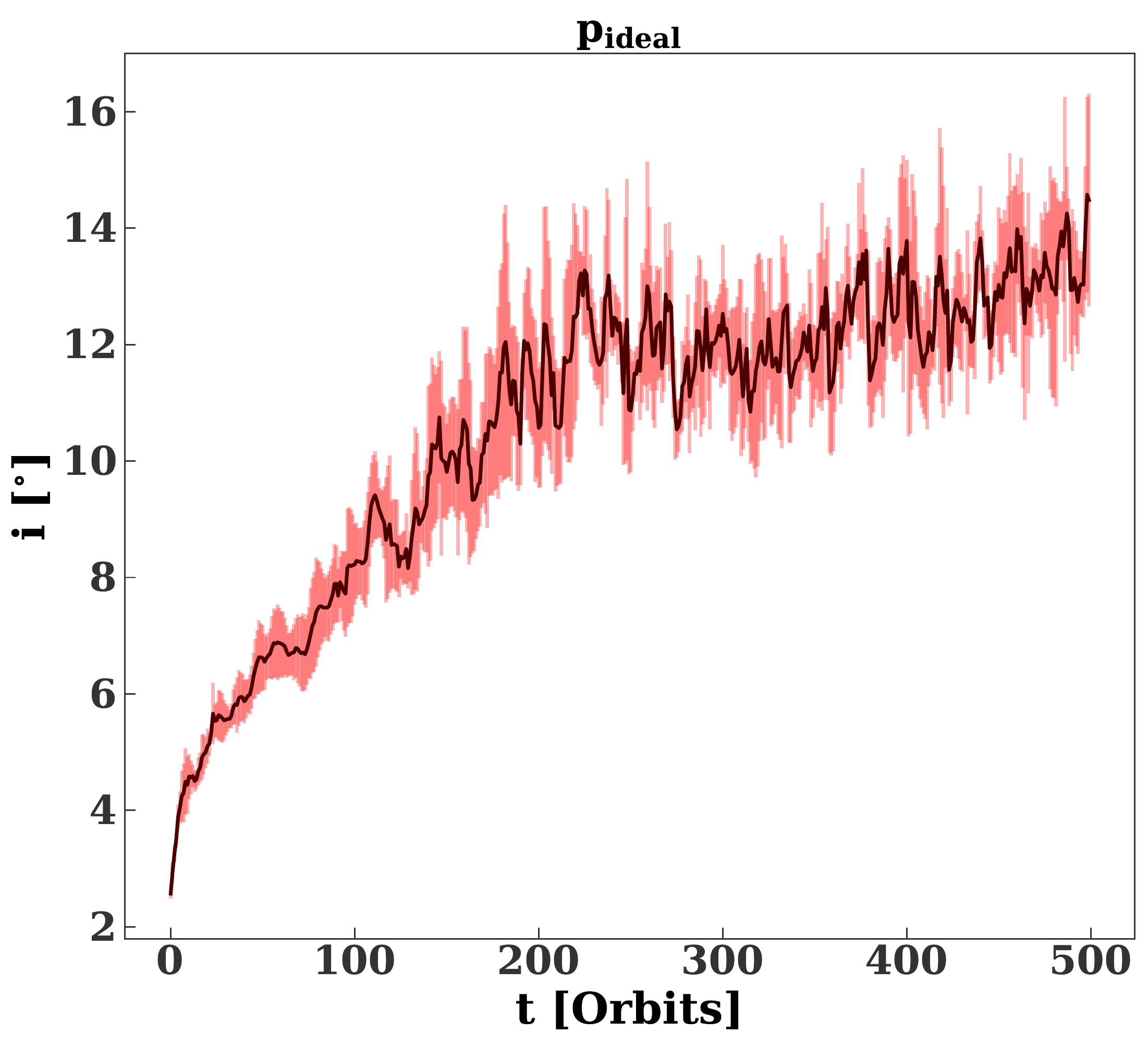}
\includegraphics[width=.45\textwidth]{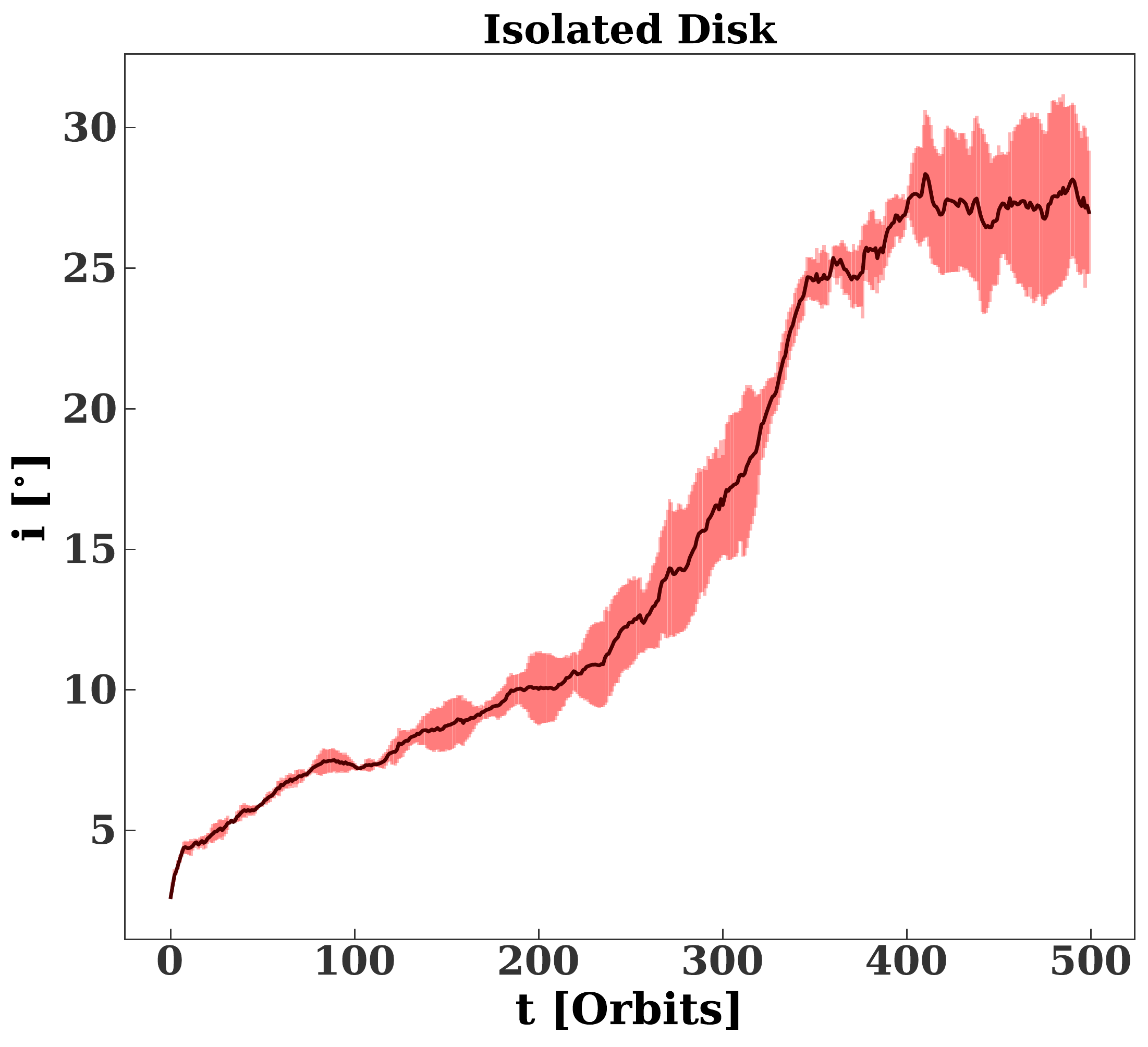}
\caption{\label{fig:incsplots} Top panel: average inclination as a function of time for $p_{\rm ideal}$ simulations ($M_p=0.5, a_p=9, i_{p}=180^{\circ}$). The line and shaded area are the average and standard deviation of four different simulations as described in Figure~\ref{fig:eccsplots}. The inclination increases for a secular time, before saturating at $\sim 14^{\circ}$. Bottom panel: average inclination as a function of time for isolated disk simulations. The inclination increases for two secular times,  consistent with \citet{foote+2020}. }}
\end{figure}

In secular dynamics, inclination and eccentricity are usually anti-correlated (due to conservation of angular momentum). Thus, the inclination of the disk significantly increases after a secular time in isolated disk simulations. This increase is damped by the perturber as shown in top panel of Figure~\ref{fig:incsplots}. 

\begin{figure}
\includegraphics[width=.45\textwidth]{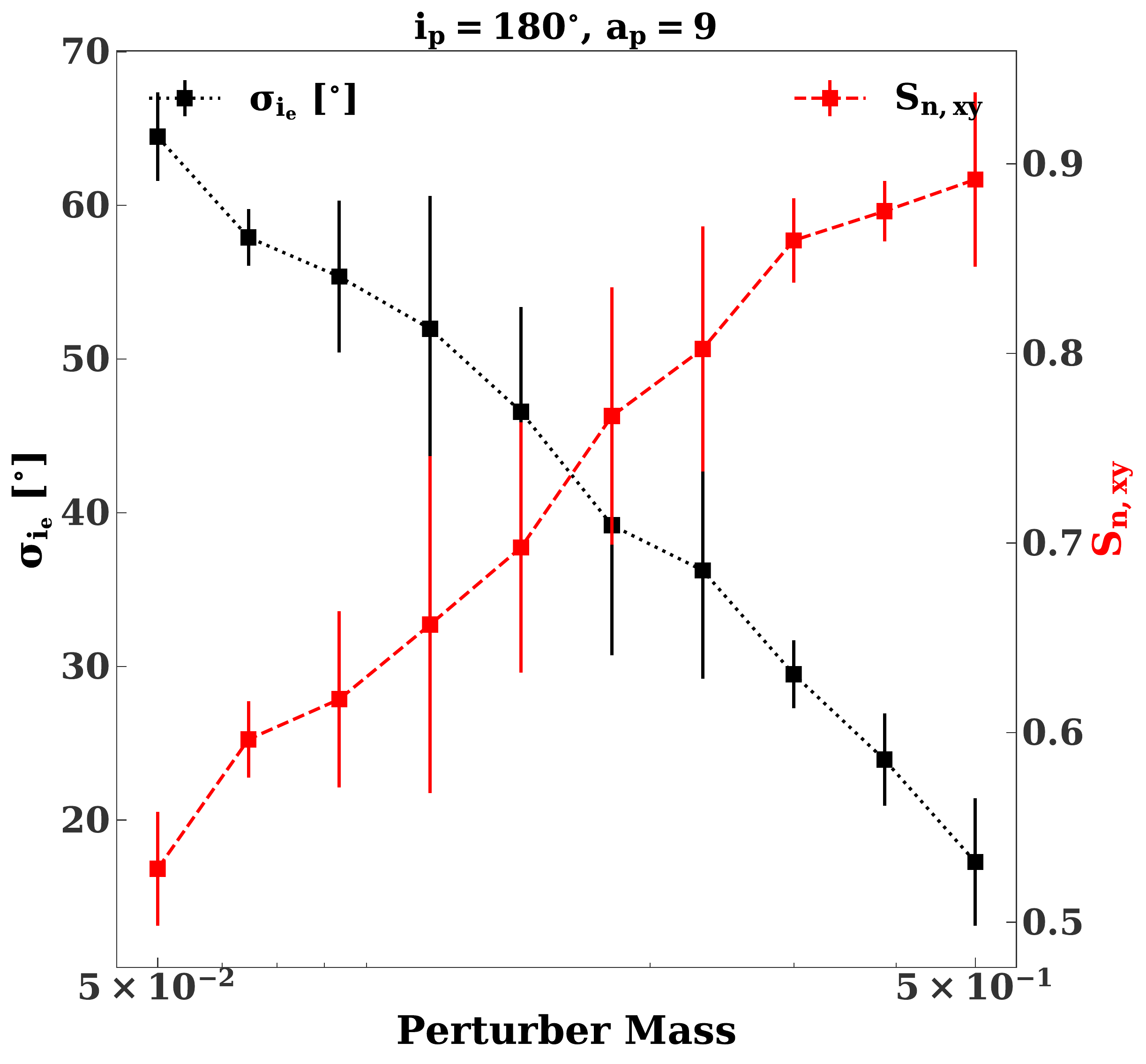}
\caption{\label{fig:newcompare}  The red, dashed line shows $S_{\rm n,xy}$, the xy-projected Rayleigh dipole statistic as a function of perturber mass, $M_p$ (for $a_p=9$ and $i_p=180^{\circ}$).  The dipole statistic varies linearly with $\log M_{p}$, and is significantly larger than the maximum expected dipole statistic for 100 isotropic vectors ($\sim$ 0.1). The dotted, black line shows the circular standard deviation of $i_e$, as a function of perturber mass, which also varies linearly with $\log (M_p)$. At large perturber masses, $S_n$ is large and $\sigma_{i_e}$ is small, indicating the disk is more aligned. Thus, the two statistics are consistent. The error bars are the standard deviation of measurements from four different simulations.
}
\end{figure}

\begin{figure}
\includegraphics[width=.45\textwidth]{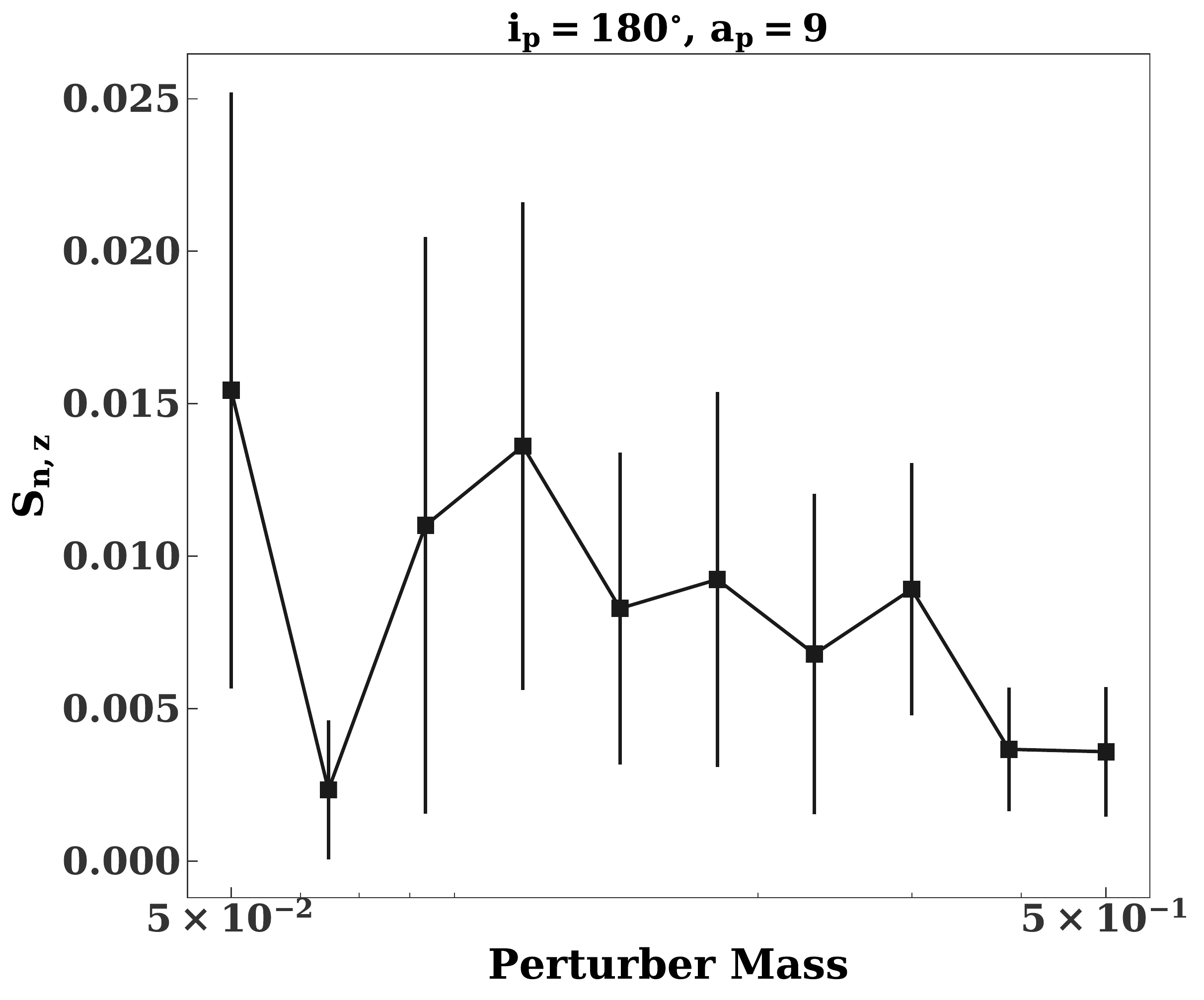}
\caption{\label{fig:Snz} The z-projected Rayleigh dipole, $S_{n,z}$, as a function of mass for perturbers that are aligned with the disk ($i_p=180^{\circ}$, $a_{p}=9$). There is no alignment of the disk's eccentricity vectors along the z-axis.  As in Figure~\ref{fig:newcompare}, the maximum expected dipole statistic for 100 isotropic vectors is $\sim$ 0.1, and all points here are below this threshold.}
\end{figure}

\begin{figure}
{
\includegraphics[width=.45\textwidth]{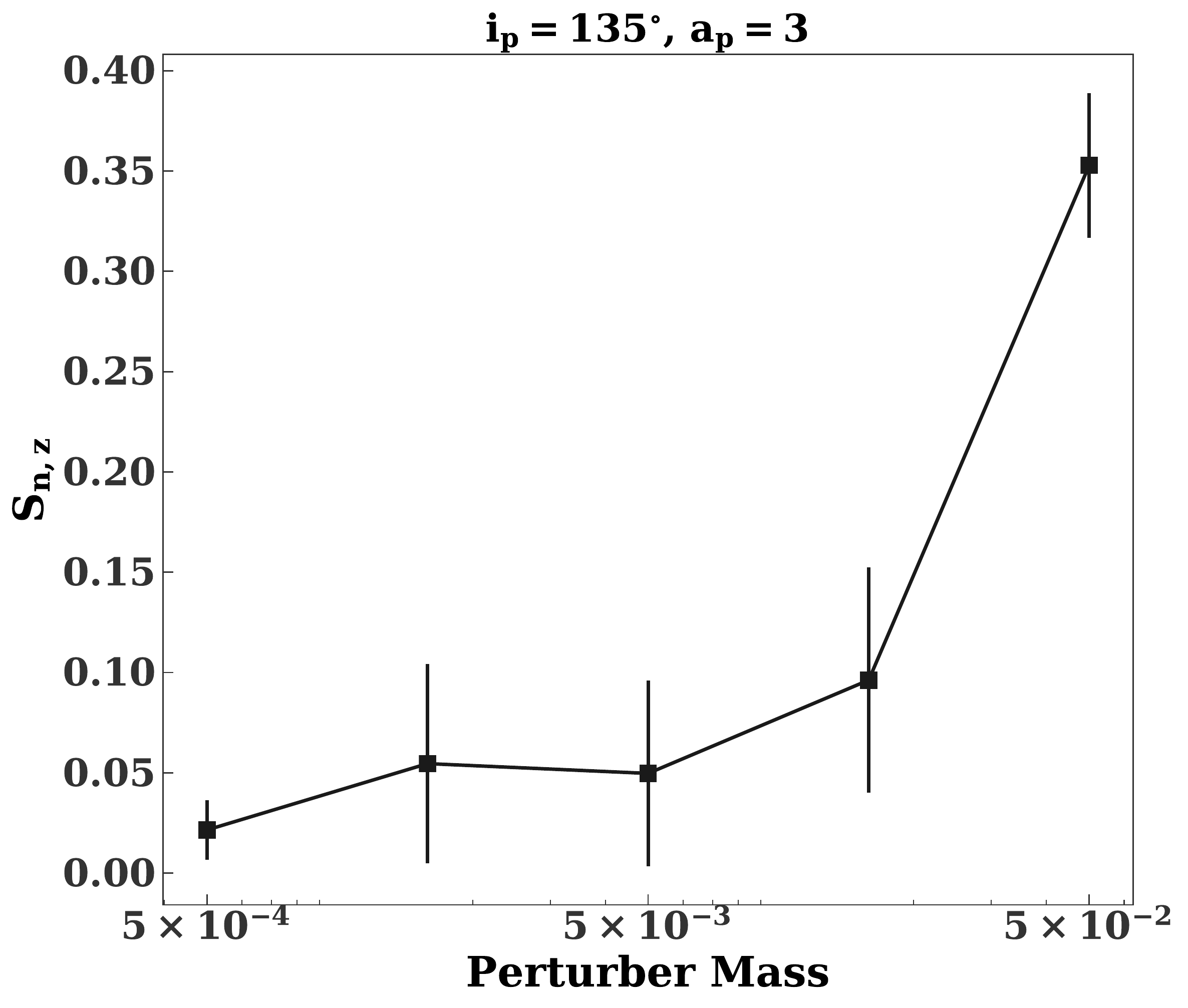}
\caption{\label{fig:Snz2} $S_{n,z}$, the z-projected Rayleigh dipole, as a function of mass for a non-coplanar perturber with semi-major axis $a_{p}=3$ and inclination $i_{p}=135^{\circ}$. 
Inclined massive perturbers can cluster eccentricity vectors along the z-axis. As in Figures~\ref{fig:newcompare} and~\ref{fig:Snz}, the maximum expected dipole statistic for 100 isotropic vectors is $\sim$ 0.1. Only the most massive perturber lies above this threshold.}
}
\end{figure}

\subsubsection{Analysis using the Rayleigh Dipole Statistic}
\label{subsubsec:Analysis using the Rayleigh Dipole Statistic}
 In \S~\ref{subsubsec:Analysis using sigma_ie} we found 
that perturbers with $M_{p}=0.5$, $a_{p}=9$, and $i_{p}=180^{\circ}$ (``$p_{\rm ideal}$'') result in the most aligned disks, with the smallest circular standard deviation of $i_{e}$ ($\sigma_{i_e}$). We now measure the clustering of eccentricity vectors in the plane of the disk using $S_{\rm n, xy}$ (the xy-projected Rayleigh dipole statistic defined \S~\ref{subec:Quantifying Disk Alignment}).

The red line in Figure~\ref{fig:newcompare} shows the xy-projected dipole statistic as a function of perturber mass (for $a_p=9, i_p=180^{\circ}$), while the black line shows $\sigma_{i_{e}}$ for the same perturbers. An increase (decrease) in the former (latter) indicates a larger disk alignment. Both statistics scale linearly with the logarithm of the perturber mass. Both show disk alignment increases with perturber mass. Overall, we conclude that the two statistics are consistent.

The Rayleigh dipole statistic can be used to measure alignment along arbitrary axes. In particular, $S_{n,z}$ measures alignment along the z-axis (see \S~\ref{subec:Quantifying Disk Alignment}).
Figure~\ref{fig:Snz} shows $S_{n,z}$ as a function of perturber mass (with $a_p=9$). As expected, there is no significant alignment of eccentricity vectors along the z-axis for coplanar perturbers. 
However, Figure~\ref{fig:Snz2} shows non-coplanar perturbers can produce alignment along the z-axis.

\subsection{A model for the disk dynamics}
\label{subsec:A model for the disk dynamics}
Certain perturbers (e.g. $p_{\rm ideal}$) reduce the angular width of eccentric disks in our simulations, indicating differential precession is suppressed. We now explore this effect in more detail. 

\begin{figure}
\includegraphics[width=.45\textwidth]{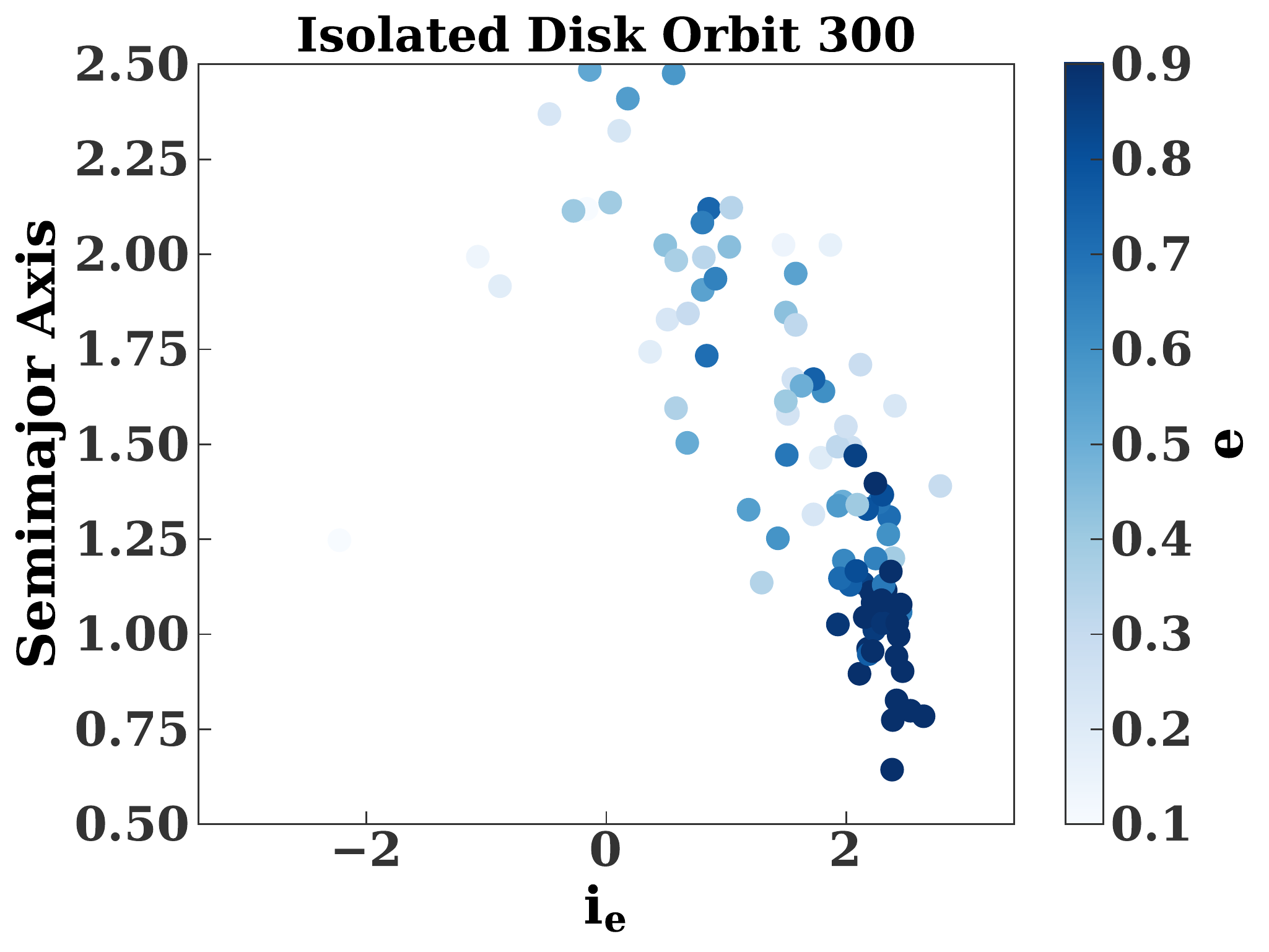}
\includegraphics[width=.45\textwidth]{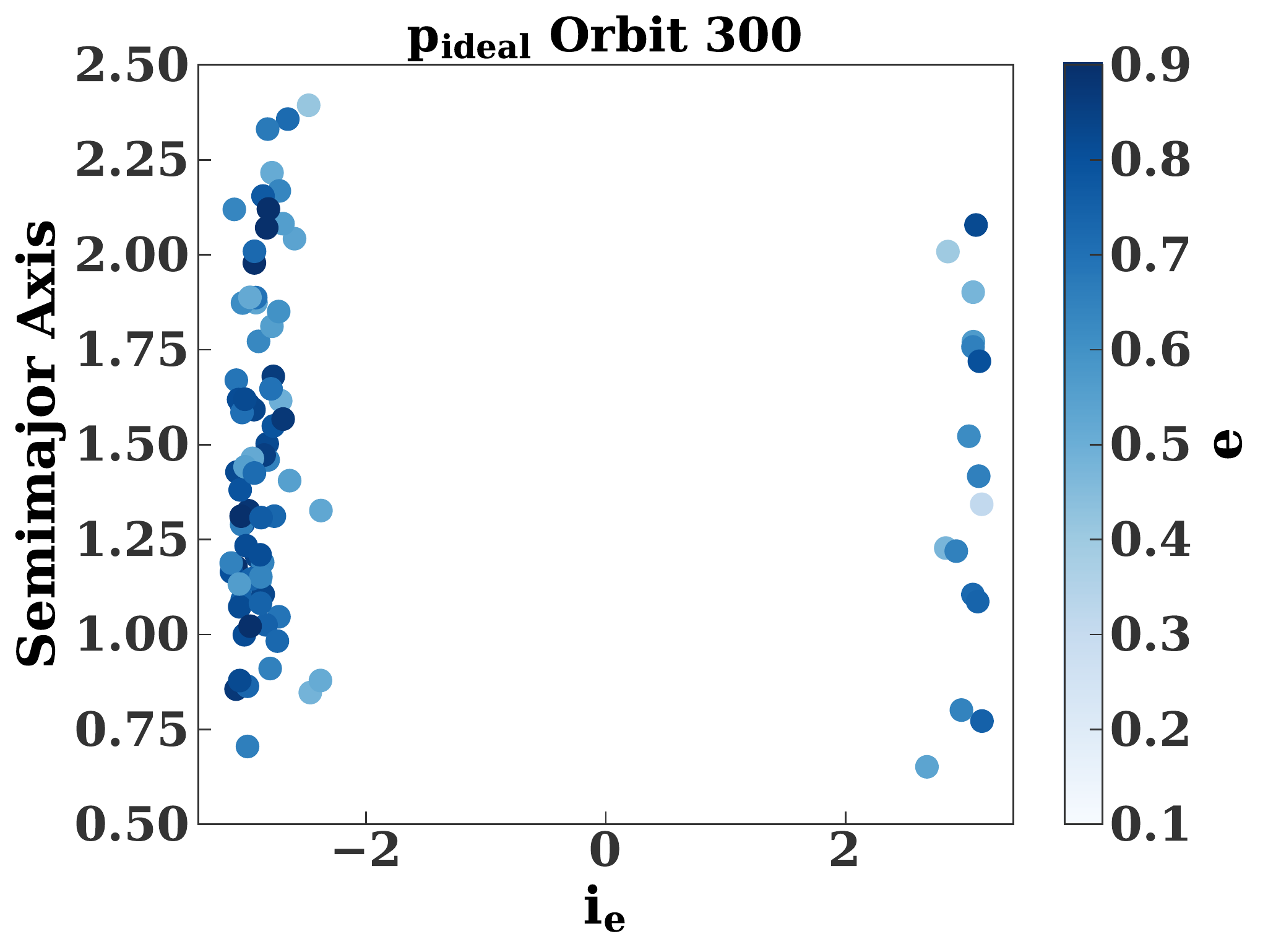}
\caption{\label{fig:Snap300_control_and_pideal}  Top Panel: Snapshot of semi-major axis as a function of $i_e$ for an isolated disk. Colors show eccentricity. The outer regions of the disk have fallen behind the inner regions, indicating differential precession. The eccentricity increases towards the inner disk. Bottom Panel:  Same as top panel, but for a $p_{\rm ideal}$ simulation ($M_p=0.5, a_p=9, i_{p}=180^{\circ}$). Here, the disk does not have any significant ``tilt,'' and there is no eccentricity gradient.}
% \end{subfigure}
\end{figure}

The top and bottom panels of Figure~\ref{fig:Snap300_control_and_pideal} show a snapshot of semi-major axis versus $i_{e}$ for a control and $p_{\rm ideal}$ simulation, respectively.
In the isolated disk case, the outer disk lags behind the inner disk. In the $p_{\rm ideal}$ simulation, this differential precession is absent. 

We can understand this behavior via a semi-analytic model.
The perturber provides an extra source of prograde precession, which can be estimated via a secular approximation as described in Appendix~\ref{sec:Appendix}. The total precession rate is the sum of the disk- and perturber-induced precession rates, viz.
\begin{equation}
    \frac{d i_{e}}{dt}=  \left.\frac{d i_{e}}{dt}\right|_{\rm disk} + \left.\frac{d i_{e}}{dt}\right|_{\rm perturber}.
    \label{eq:precTot}
\end{equation}
The first term can be estimated from the simulations of an isolated disk. In particular, we divide stars into five evenly-spaced semi-major axis bins, and compute the change in the mean $i_{e}$ in each bin over ten orbits. We measure the precession rate at 100 orbits, before the disk has undergone significant secular evolution.  If differential precession is damped at early times, the disk will simply precess as a solid body. 

\begin{figure}
{
\includegraphics[width=.45\textwidth]{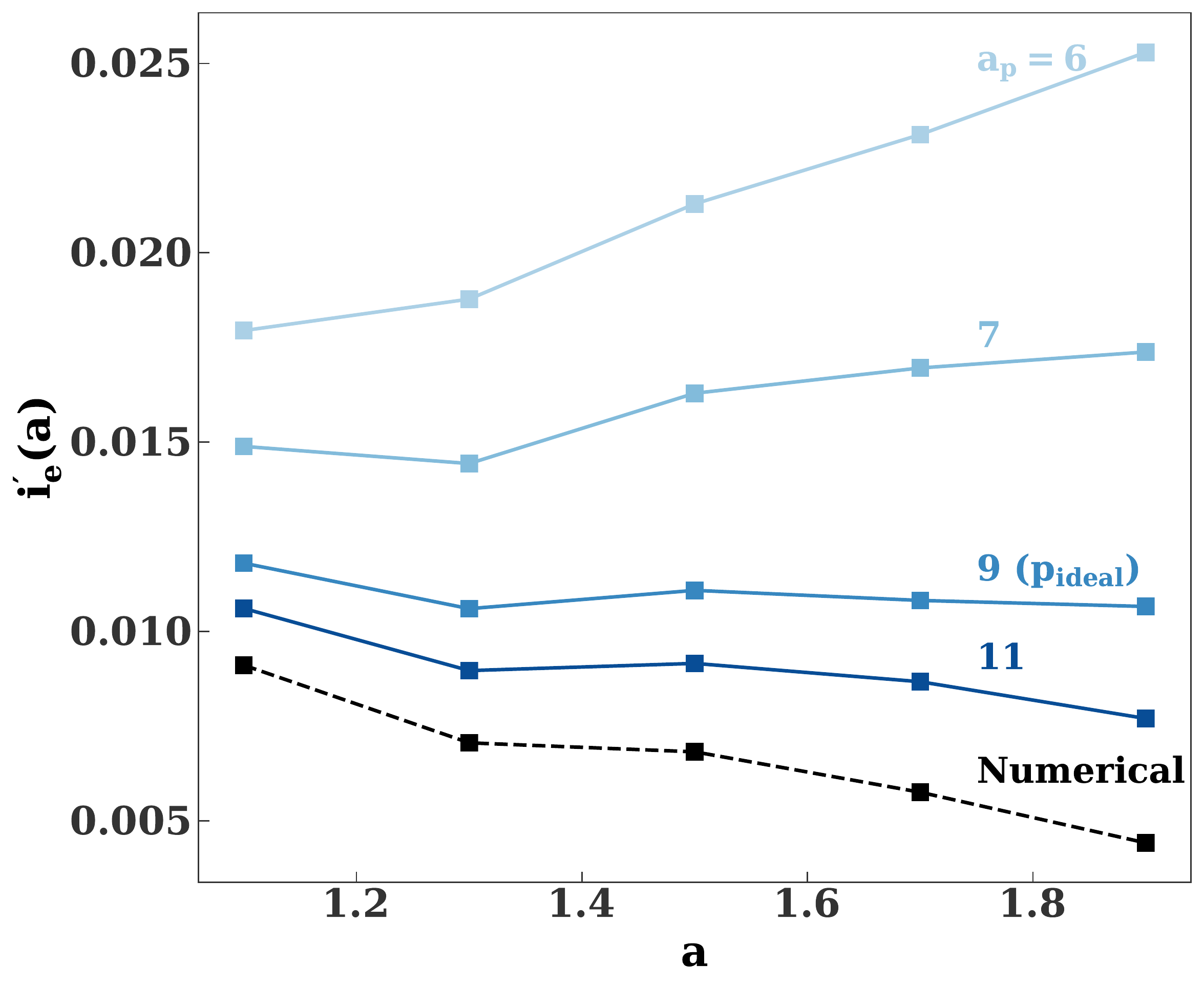}
\caption{\label{fig:proto_numeric_analytic} Precession rate as a function of disk semi-major axis for isolated  (\emph{black line}) and perturbed (\emph{colored lines}) eccentric nuclear disks. The different line colors correspond to different perturber semi-major axes (the perturber mass is 0.5). In each case, we add the analytic perturber-induced precession rate to the  precession rate measured in isolated disk simulations. The total precession rate is flattest near $p_{\rm ideal}$ ($M_p=0.5, a_p=9, i_{p}=180^{\circ}$).}
}
\end{figure}

\begin{figure}
{
\includegraphics[width=.48\textwidth]{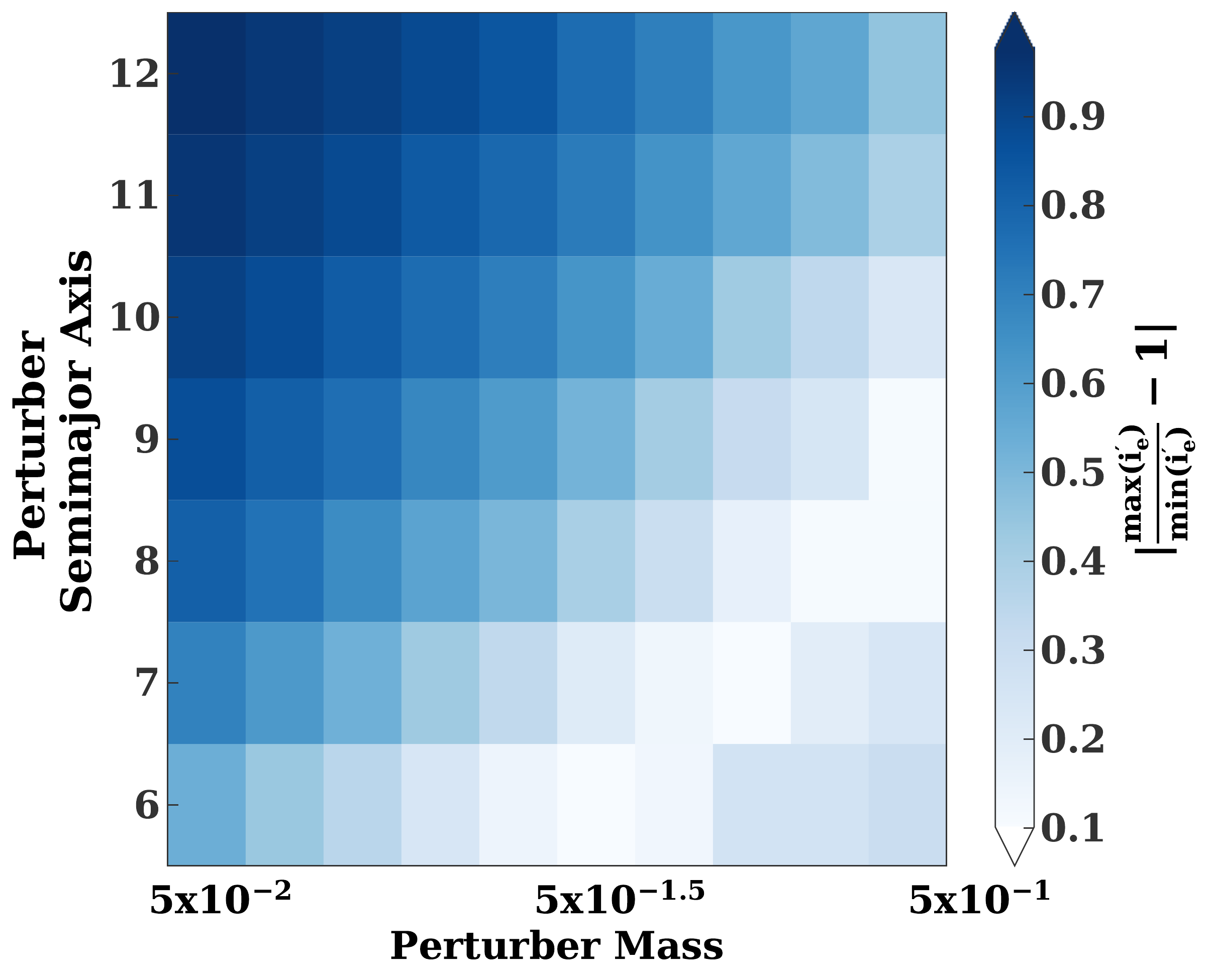}
\caption{\label{fig:analyticnumeric} Fractional difference between the maximum and minimum precession rate in the disk as a function of perturber mass and semi-major axis, according to equation~\eqref{eq:precTot} in \S~\ref{subsec:A model for the disk dynamics}. The behavior is qualitatively similar to that in Figure~\ref{fig:ie4}. Specifically, regions with low differential precession have strong disk alignment.}}
\end{figure}

Figure~\ref{fig:proto_numeric_analytic} shows the total precession rate according to equation~\eqref{eq:precTot} as a function of disk and perturber semi-major axis for the largest mass perturbers in our grid.  As expected, the $p_{\rm ideal}$ perturber results in the smallest differential precession rate (i.e. the total precession rate is flat with semi-major axis). In fact, the model approximately reproduces the dependence of disk alignment on perturber mass and semi-major axis, as shown in Figure~\ref{fig:analyticnumeric}. Perturbers with the smallest differential precession in Figure~\ref{fig:analyticnumeric} correspond to perturbers with the smallest $\sigma_{i_{e}}$ in Figure~\ref{fig:ie4}. 

To summarize, in certain regions of parameter space (e.g. $p_{\rm ideal}$) perturbers shut down differential precession in an eccentric disk. Although we have only considered retrograde, coplanar perturbers, inclinations above $160^{\circ}$ produce similar results.

With a lack of differential precession, many of the secular dynamical effects that are present in an isolated disk are also suppressed. For example, in an isolated disk with (initially) uniform eccentricity, orbits near the inner edge of the disk would precess ahead of the outer disk. These orbits would then be torqued to a lower angular momentum (higher eccentricity), and slow down so that they are more aligned with the outer disk. Therefore, an eccentricity gradient develops in an isolated disk. However, with a perturber a nearly uniform eccentricity profile can be a stable configuration, which preccesses as a solid body. Thus, the eccentricity gradient is suppressed, as shown in Figure~\ref{fig:Snap300_control_and_pideal}. The suppression of the eccentricity gradient also suppresses TDEs.

\subsection{TDEs}
\label{subsec:TDEs}
In an isolated disk, orbits oscillate about the disk, simultaneously changing their eccentricities (see Figure 5 in \citealt{madigan+2018}). As their eccentricities oscillate, stars may experience a TDE. 
Perturbers can suppress these secular oscillations, and the resulting TDEs. Furthermore, in an isolated eccentric disk there is an eccentricity gradient, and most TDEs come from the inner edge of the disk where the eccentricity is highest. In some perturber simulations (e.g. $p_{\rm ideal}$) the eccentricity gradient is no longer present, and the eccentricity at the inner edge of the disk is smaller than in the isolated case -- further suppressing TDEs.

\begin{figure}
\includegraphics[width=.45\textwidth]{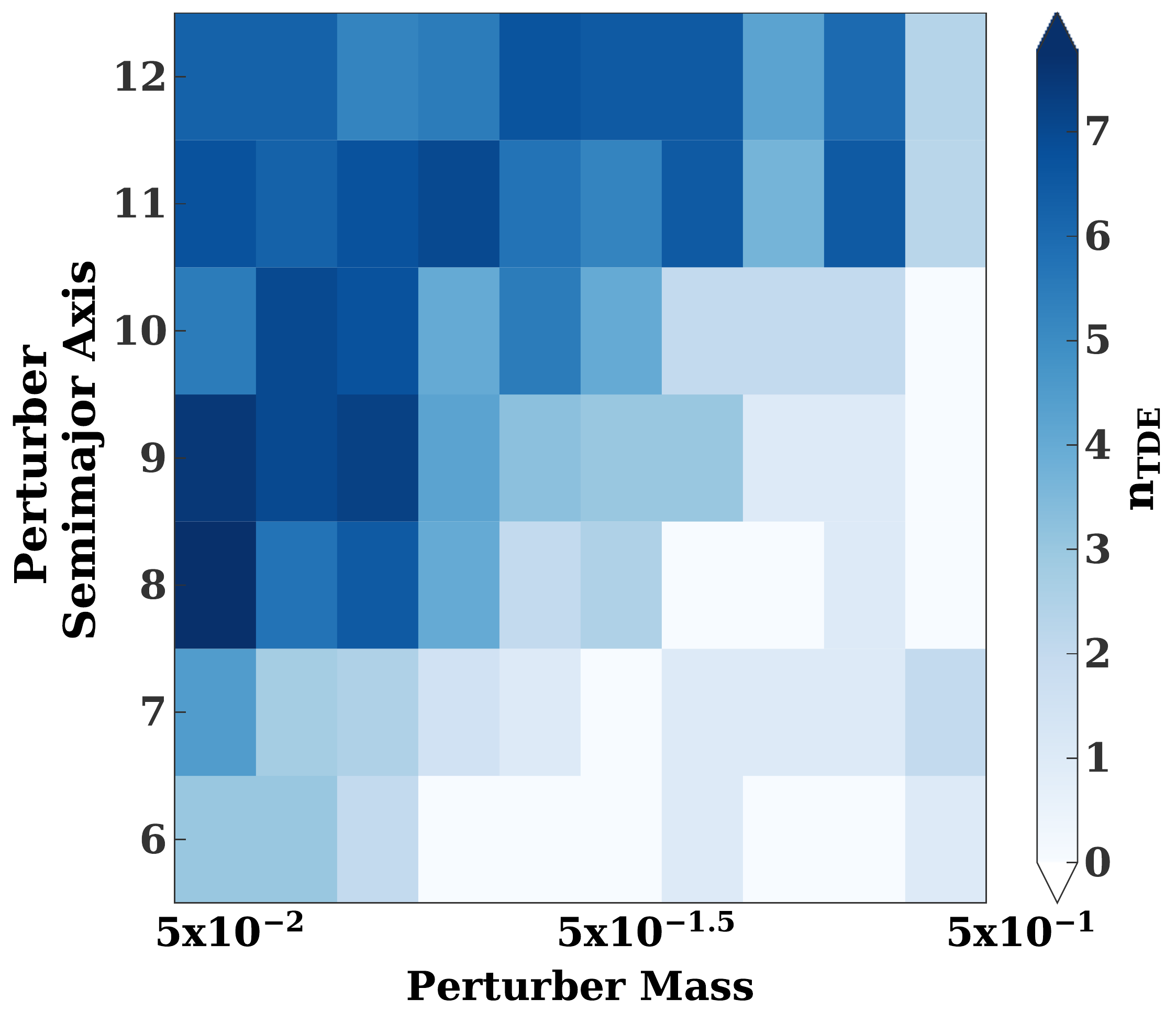}
\caption{\label{fig:tdeplot} Number of tidal disruption events (TDEs) as a function of perturber mass and semi-major axis for retrograde perturbers. In certain regions of parameter space, perturbers suppress secular oscillations in the disk, reducing the TDE rate. Some simulations have no TDEs. }
\end{figure}

Figure~\ref{fig:tdeplot} shows the number of TDEs as a function perturber mass and semi-major axis for the simulations in Table~\ref{tab:params2}. In comparison, there are $7\pm 1$ TDEs in our isolated disk simulations.\footnote{The number of TDEs fluctuates in simulations due to Poisson noise. This fluctuation will be comparable to the square-root of the number of TDEs.} The variation in the number of TDEs mirrors the variation in disk alignment in Figure~\ref{fig:ie4}: the smaller the disk spread, the fewer TDEs occur. In some cases (e.g. $p_{\rm ideal}$) there are zero TDEs. 

\subsection{Prograde vs. retrograde orbits}
\label{subsec:Prograde vs. retrograde orbits}
So far, we have used the secular approximation in Appendix~\ref{sec:Appendix} to estimate the perturber-induced precession rate in the disk. This approximation neglects non-secular effects, and predicts no differences between prograde and retrograde perturbers. 

\begin{figure}
{
\includegraphics[width=.45\textwidth]{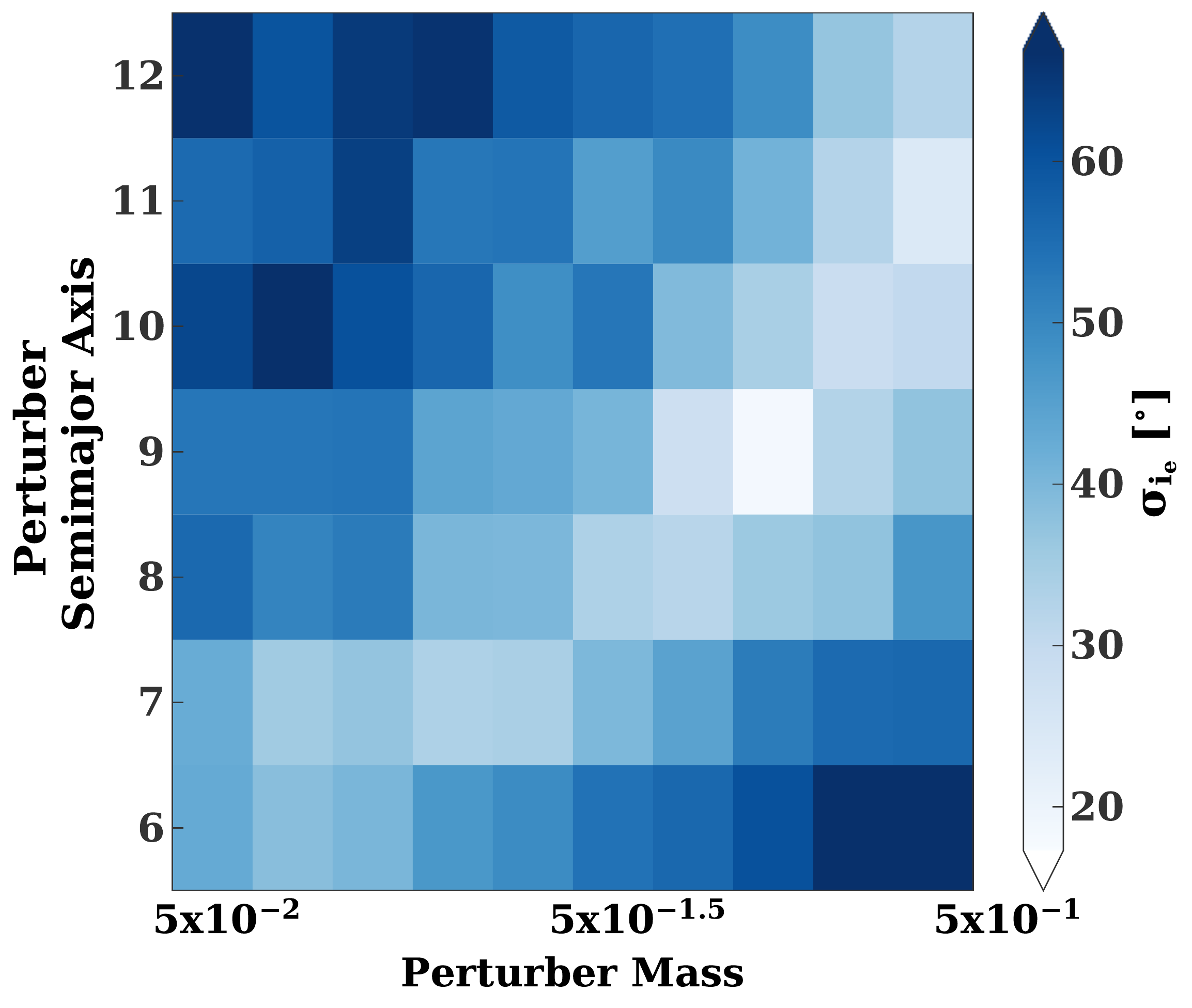}
\caption{\label{fig:ie5}  Circular standard deviation of $i_e$ as a function of perturber mass and semi-major axis for prograde perturbers (with $i_p=0^{\circ}$). Our secular model predicts that prograde and retrograde perturbers should produce equivalent effects, yet compared to Figure~\ref{fig:ie4} maximal alignment occurs at lower mass (albeit at the same semi-major axis).
}
}
\end{figure}

To test this approximation, we replicate Figure~\ref{fig:ie4} with a perturber at $i_p = 0^{\circ}$ instead of $180^{\circ}$.  The results are shown in Figure~\ref{fig:ie5}. The conditions for maximal alignment are shifted to lower masses, meaning the secular approximation provides an incomplete picture of the dynamics. 

We can improve our model by measuring the perturber-induced precession rate via direct three-body simulations, as described in Appendix~\ref{sec:AppendixNum}. The secular model most closely matches these numerical integrations for small mass, large semi-major axis, and retrograde perturbers. Non-secular effects are more prominent for prograde perturbers as the relative velocities are smaller at close separations.

\begin{figure}
\includegraphics[width=.48\textwidth]{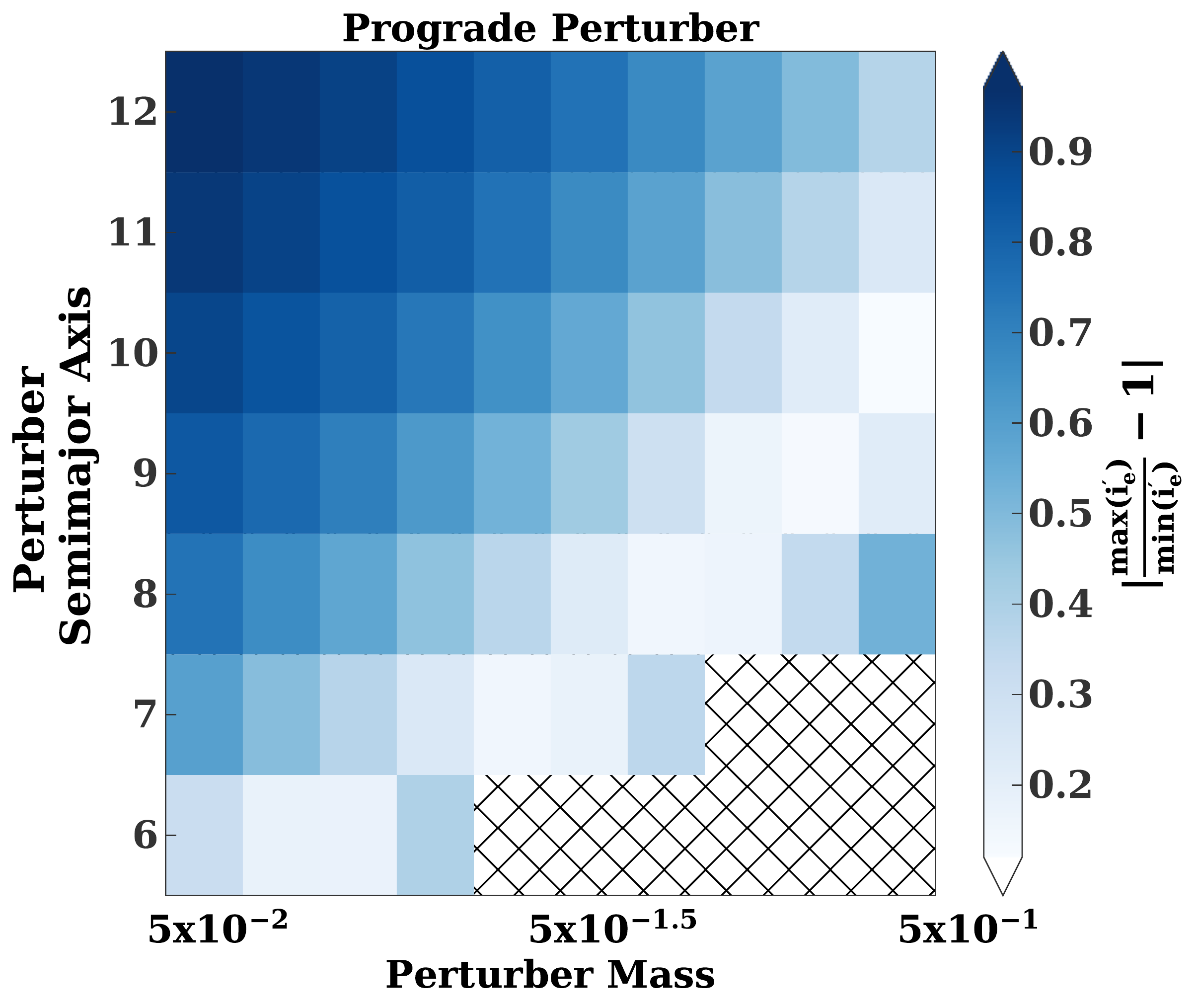}
\includegraphics[width=.48\textwidth]{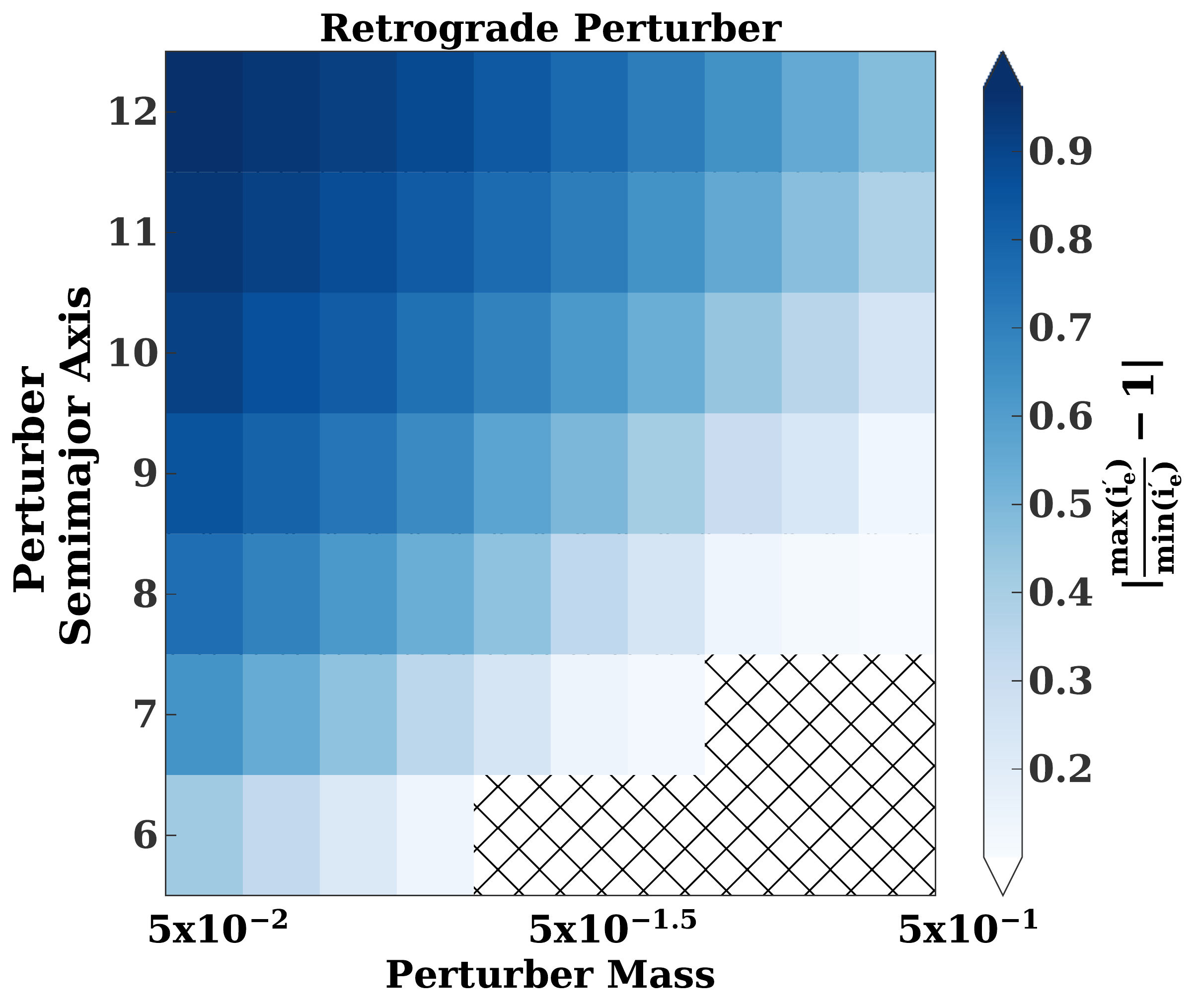}
\caption{\label{fig:precs_all} Same as Figure~\ref{fig:analyticnumeric}, except the perturber-induced precession rate is calculated via three-body integrations, as described in Appendix~\ref{sec:AppendixNum}. The parameters that minimize differential precession are somewhat different for prograde (top panel) and retrograde (bottom panel) perturbers. Similarly, the parameters that maximize disk alignment are somewhat different for prograde and retrograde perturbers (as previously shown in Figures~\ref{fig:ie4} and~\ref{fig:ie5}).  Note that we do not show the differential precession rate for the cross-hatched region in the bottom, right corner, as these perturbers induce highly scattered orbits (which no longer have a well-defined precession rate).
}
\end{figure}

Figure~\ref{fig:precs_all} shows effects of the above modification on the total precession rate in the disk, as estimated by equation~\eqref{eq:precTot}. The top and bottom panels qualitatively reproduce the behavior of disk alignment in simulations with prograde (see Figure~\ref{fig:ie5}) and retrograde perturbers (see Figure~\ref{fig:ie4}) respectively. 

\section{Discussion}
\label{sec:Discussion}
\subsection{SMBH inspiral timescale}
\label{subsec:SMBH inspiral timescale}
In this section, we place our results in a broader astrophysical context. In particular, we discuss the differences between our simulations and more realistic eccentric disk systems. For example, our simulations have far fewer particles than a realistic disk. They also neglect hardening of the SMBH binary due to scatterings with background stars. We review the timescales for SMBH binary inspirals, and evaluate the applicability of this assumption.

\begin{figure}
    \includegraphics[width=.45\textwidth]{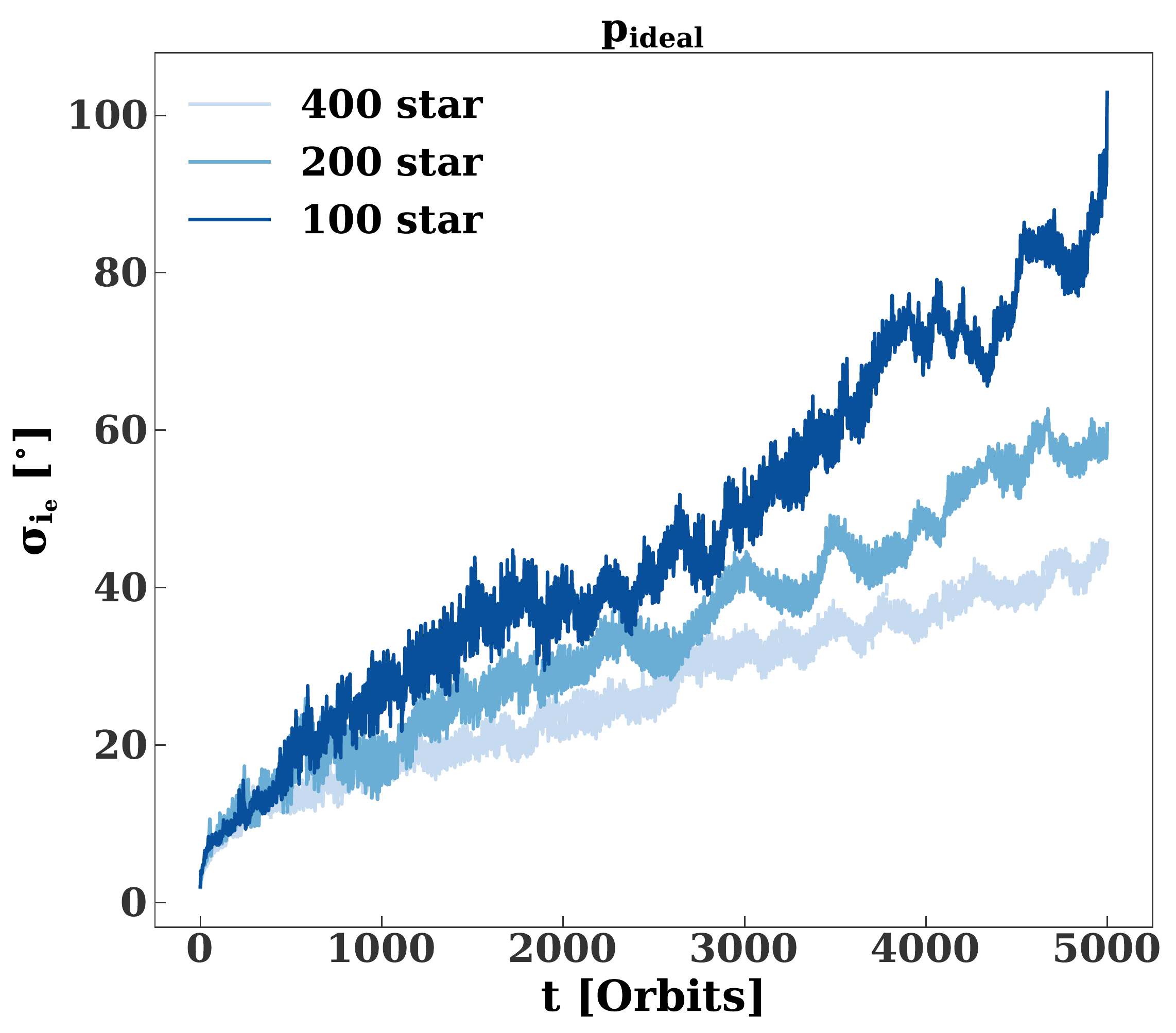}
    \caption{Circular standard deviation of $i_e$, as a function of time for $p_{\rm ideal}$ ($M_p=0.5, a_p=9, i_{p}=180^{\circ}$) simulations with 100 (dark blue line), 200 (blue line), and 400 (light blue line) stars (with the same disk mass). For fixed disk mass, the two-body relaxation time is approximately proportional to the number of stars. Therefore, the disk spreads out more in simulations with fewer stars, and the angular spread of the simulated disks would be unreliable at late times. 
    }
    \label{fig:200star100star}
\end{figure}

We simulate eccentric disks with $100$ stars, although realistic disks would have stars many orders of magnitude more.  This means two-body relaxation is artificially strong in our simulations. Two-body relaxation will cause the disks in our simulations to spread out much more rapidly than a realistic systems.
This effect is shown for $p_{\rm ideal}$ in Figure~\ref{fig:200star100star}, which compares the evolution of disk alignment in 100, 200, and 400 star disks (with the same disk mass). Although the results diverge at late times, the evolution is similar before $\sim$ 500 orbits, suggesting that two-body relaxation does not play a significant role before this time.

We now turn to the feasibility of a fixed binary orbit based on previous work on SMBH binaries.
In a seminal paper, \citet{begelman+1980}, described the formation and evolution of SMBH binaries in galaxy mergers. The formation process can be divided into several stages, viz.
\begin{enumerate}
    \item The SMBHs sink towards the center of the merger remnant via dynamical friction, until they form a hard binary. 
    \item The SMBHs continue to harden via three-body scatterings with the surrounding stellar population.
    \item Gravitational wave emission brings the SMBH binary to merger.
\end{enumerate}
The dynamical friction timescale for a bare SMBH in an isothermal sphere with velocity dispersion $\sigma$ is (cf \citealt{binney&tremaine1987})
\begin{equation}
    t_{\rm df}=\frac{19}{\log \Lambda} \left(\frac{\sigma}{200 {\rm \,\,km\,\, s^{-1}}} \right) \left(\frac{10^8 M_{\odot}}{M_p}\right)^{-1} \left(\frac{r_i}{5 {\rm kpc}}\right)^{2} {\rm Gyr},
    \label{eq:tdf1}
\end{equation}
where $r_i$ and $M_p$ are the starting radius and mass of the SMBH; $\log \Lambda \approx \log (2 r_i \sigma^2/G M_p)$ is the Coulomb logarithm.
In fact, the dynamical friction time is likely considerably shorter, considering each SMBH will be surrounded by stars from its host galaxy. 
As described in \citet{dosopoulou&antonini2018}, the total surviving mass around the SMBH would be set by tidal truncation. Following their prescriptions (see their equation 57), the inspiral timescale for a $10^{8}$ ($10^7$) $M_{\odot}$ SMBH in a $\sigma=200$ km s$^{-1}$ isothermal sphere is $\sim 2 \times 10^7$ ($4\times 10^8$) yr. Note we have assumed (i) the SMBH is initially surrounded by an isothermal bulge with $10^3$ times its mass, and (ii) the velocity dispersion of this bulge is $\sigma_s= 200$ km s$^{-1} (m/10^8 M_{\odot})^{1/4.3}$, as suggested by the $M-\sigma$ relation \citep{gultekin+2009, kormendy&ho2013}.

The dynamical friction inspiral ends once the two SMBHs find each other and form a hard binary  \citep{binney&tremaine1987}. There is some ambiguity in the definition of a hard binary \citep{merritt&szell2006}. One estimate is 
\begin{align}
    a_h&=\frac{G m_{\rm bin}}{4 \sigma^2} \frac{q}{(1+q)^2}\nonumber\\
      &=2.7 {\rm pc} \left(\frac{m_{\rm bin}}{10^8 M_{\odot}}\right) \left(\frac{\sigma}{200 {\rm \,\,km\,\, s^{-1}}}\right)^{-2} \frac{q}{(1+q)^2},
      \label{eq:ah1}
\end{align}
where q is ratio between the secondary and primary mass. For $q=0.5$, $m_{\rm bin}=10^8 M_{\odot}$, and $\sigma=200$ km s$^{-1}$, $a_h=0.7$ pc. However, the velocity dispersion is poorly defined for a non-isothermal stellar population. Another definition, which avoids this ambiguity is
\begin{align}
    a_h&= \frac{q}{4 (1+q)^2} r_{\rm inf}
    \label{eq:ah2}
\end{align}
where $r_{\rm inf}$ is the radius enclosing twice the mass of the binary \citep{merritt&wang2005, merritt&szell2006}. For an M31-like density profile\footnote{Specifically we use the Einasto law fits to the M31 bulge and nucleus from Table 2 in \citet{tamm+2012}.}, and a $10^8 M_{\odot}$ primary, $r_{\rm inf} \approx 70 \left(\frac{1+q}{2}\right)^{1/(3-\gamma)}$ pc, where $\gamma \approx 0.7$ is the slope of the stellar density profile. Thus,
\begin{align}
    a_{h, M31}\approx 13 {\rm pc} \frac{q}{(1+q)^{1.56}}.
\end{align}
This is 3.4 pc for $q=0.5$, which corresponds to $p_{\rm ideal}$ for a disk with inner edge at 0.4 pc. (For comparison the inner edge of the M31 disk is somewhat larger. Its surface density falls off steeply inside of 2 pc; \citealt{brown&magorrian2013}.)

\begin{figure}
    \includegraphics[width=8.5cm]{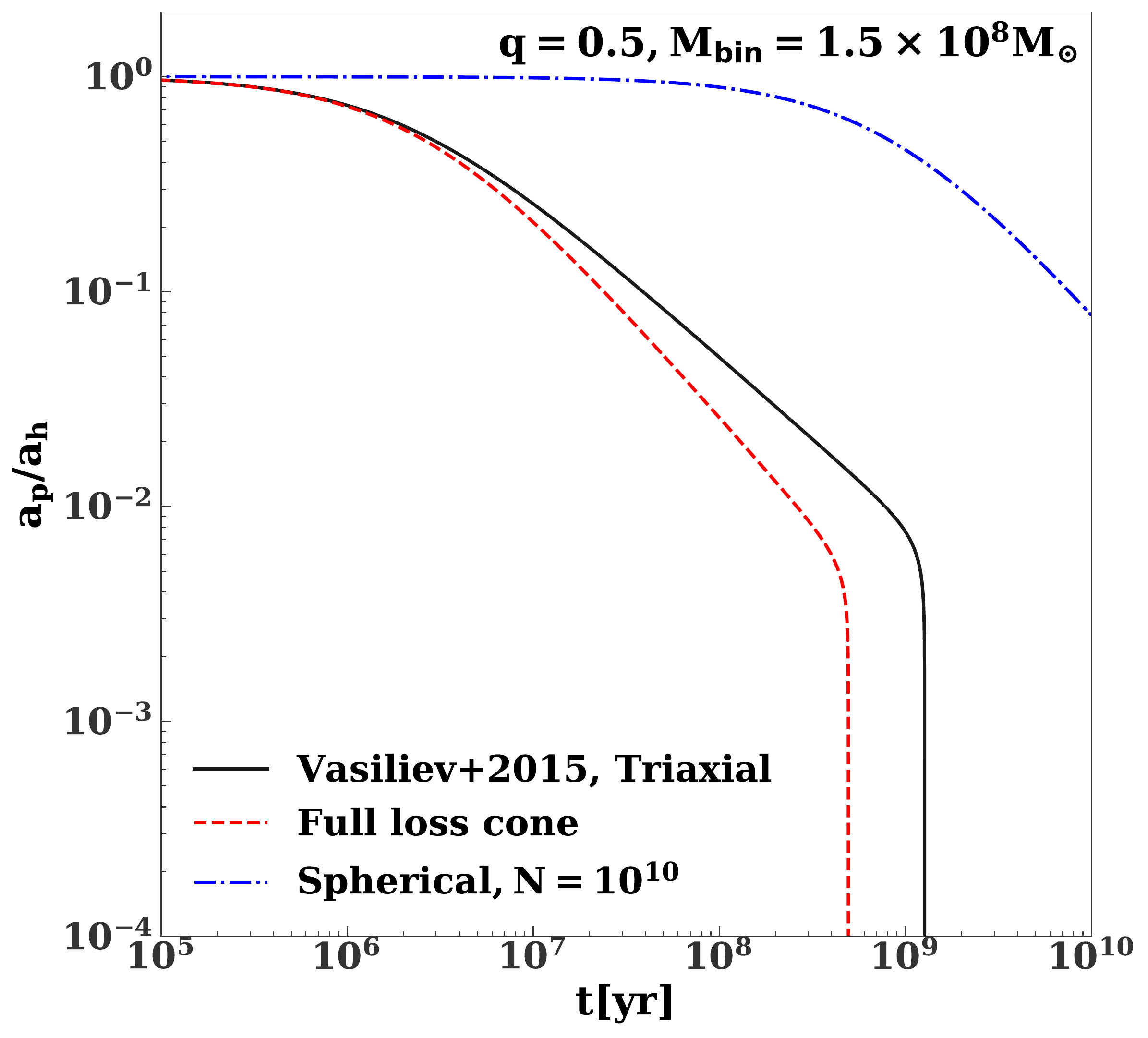}
    \caption{Semi-major axis as a function of time for a hard SMBH binary in a triaxial nucleus, based on \citetalias{vasiliev+2015} (solid, black line).  Up to a few$\times 10^9$ years, the binary gradually loses energy to slingshot scatterings of stars. At this point, gravitational wave emission dominates the evolution of the binary, and rapidly brings it to coalescence.  The evolution is similar to that of a binary hardening at the full loss cone rate (dashed, red line). In a spherical nucleus, however, the loss cone will be far from full. The blue, dashed-dotted line shows the evolution of the binary semi-major axis, assuming equation~\ref{eq:shard} for the hardening rate due to three-body encounters in a spherical nucleus (with $N=10^{10}$).}
    \label{fig:vasExample}
\end{figure}

At this stage, the SMBH binary can continue to harden as nearby stars ``slingshot'' off of it, as they come within the semi-major axis of the binary \citep{saslaw+1974}.
Figure~\ref{fig:vasExample} shows the semi-major axis evolution for a $10^8 -5\times 10^7 M_{\odot}$ SMBH binary as a function of time in a triaxial nucleus according to equation 23 in \citetalias{vasiliev+2015}. The semi-major axis of the binary, $a_p$, evolves according to 

\begin{align}
    \frac{d(1/a_p)}{dt}= S_{*,h} \left(\frac{a_p}{a_h}\right)^{\nu}+S_{\rm gw}\nonumber\\
    S_{*,h}= A (G M_{\rm bin})^{1/2} r_{\rm inf}^{-1/2}
\end{align}
where the the first and second term on the right-hand side are the hardening rates due to 
three-body scatterings, and gravitational wave emission respectively; $S_{*,h}$ is the hardening rate at $a_h$, which depends on the influence radius ($r_{\rm inf}$) and the binary mass ($M_{\rm bin}$) (see also \citealt{sesana&khan2015}).\footnote{We take $\nu=0.3$, $A=4$, and $r_{\rm inf}=70$ pc.  We neglect the eccentricity evolution of the binary.} The binary semi-major axis approximately decreases as $t^{-0.7}$ after 1 Myr. In this case, the rate of binary hardening is close to the maximal full loss cone rate (the dashed, red line in this figure). 

In a spherical nucleus loss cone refilling would occur via two-body relaxation and would be much slower. 
In direct $N$-body simulations, \citetalias{vasiliev+2015} find that the hardening rate of an SMBH binary in a spherical nucleus is 
\begin{equation}
    S\equiv \frac{d (1/a_{p})}{d t} \sim S_{\rm full} \left(\frac{N}{10^5}\right)^{-1/2},
    \label{eq:shard}
\end{equation}
where $N$ is the total number of stars in their galaxy, which is one hundred times the mass of the binary. Thus, $N\sim 10^{10}$ for $M_{\rm bin}=10^8 M_{\odot}$. Therefore, as shown by the blue, dashed-dotted line in Figure~\ref{fig:vasExample},
the binary evolution is much slower for spherical nuclei.
In fact, the hardening rate may fall off even more steeply for very large $N$, asymptotically approaching $N^{-1}$.\footnote{The $N$-scaling of equation~\eqref{eq:shard} is due to (i) changes in the two-body relaxation time (which is proportional to $N/\log(N)$) and (ii) changes in the Brownian ``wandering radius'' of the binary (which is proportional to $N^{-1/2}$). However, for very large $N$, the Brownian motion would become negligible, and the hardening rate would simply scale as the inverse of the two-body relaxation time.} 

We conclude that, under certain conditions, configurations similar to $p_{\rm ideal}$ can survive for $\sim 10^{8}$ years in an M31-like nucleus (with a somewhat more compact eccentric disk). 

\section{Summary}
\label{sec:Summary}
In this paper, we present the first study of an eccentric nuclear disk within a galaxy merger scenario. Specifically, we consider the gravitational effects of a secondary SMBH on a dynamically cold disk (i.e. with a small spread in initial inclination).  Here, we present a summary of our main results.
\begin{enumerate}
    \item \textit{Disk Alignment:} We find that a perturbing SMBH can increase the apsidal alignment of an eccentric nuclear disk, by suppressing differential precession.
    \item \textit{The Eccentricity Profile of Stars:} A negative eccentricity gradient must be present for an isolated disk to stably precess (this statement is dependent on other disk properties like the surface density of disk, however the surface density profile would have to be finely tuned to prevent an eccentricity gradient).  However, with a second SMBH, a nearly uniform eccentricity profile can be stable against precession. The eccentricity profile is correlated with disk alignment: more aligned disks have a uniform eccentricity profile and more axisymmetric disks have a negative eccentricity gradient. The average eccentricity increases with disk alignment.
    \item \textit{TDE rates:} Isolated eccentric nuclear disks can have  enormous tidal disruption event (TDE) rates $10^{2}-10^{3}$ times higher than isotropic clusters.  However, secondary SMBHs that suppress the eccentricity gradient also suppress TDEs. In an isolated disk the orbits oscillate about the disk, simultaneously changing their eccentricities~\citep{madigan+2018}. When a star's eccentricity oscillates, it may experience a TDE as it passes through pericenter during a high eccentricity phase of oscillation. Since the secondary SMBH suppresses the secular oscillations of orbits, it also suppresses TDEs.
\end{enumerate}

These effects are strongest for systems in which the secondary orbits at approximately ten times the inner disk edge. This is comparable to the radius where an SMBH binary inspiral might stall for a 0.4 pc scale disk (somewhat smaller than M31's disk; see \S~\ref{sec:Discussion}) . However, the optimal separation must depend on other parameters we have not explored in any detail -- the extent and mass of the disk, eccentricity, inclination, etc.

In this work we have focused on the effects of a second SMBH on the dynamics of an existing eccentric disk, and, in particular, its effects on the disk's TDE rate. However, galaxy mergers and SMBH binaries can affect the TDE rate in other ways. 
First, SMBH binaries may affect the overall rate of TDEs via Kozai-Lidov oscillations \citep{ivanov+2005}, chaotic three-body scatterings \citep{chen+2009, chen+2011}, or by triggering bursts of nuclear star formation \citep{pfister+2020}.

Our simulations in this paper place perturbing black holes on circular orbits. In reality, the eccentricity can be excited by stellar encounters  \citep[and secular gravitational interactions within rotating clusters;][]{Madigan2012} and damped by gravitation radiation (at very small separations). However, the growth of eccentricity from stellar encounters is slow and in fact goes to zero for circular orbits \citep{quinlan1996, sesana+2006, khan+2013, vasiliev+2015}.

Finally, we note that recent simulations have found that a major merger (with a mass ratio of 4:1 or less), could reproduce many of the the large scale properties of the M31 galaxy \citep{hammer+2018}. They infer that the two nuclei coalesced 1.7-3.1 Gyr ago. What about the black holes? The observed state of M31's eccentric nuclear disk may give some clues. It is itself an old structure, with similar colors to the $>4$ Gyr old bulge in M31 \citep{olsen+2006, saglia+2010, lockhart+2018}. It also has an eccentricity gradient  \citep{brown&magorrian2013}, as expected for an isolated disk. This latter observation suggests that either the disk is in fact somewhat younger than the bulge, and formed after the two black holes coalesced, or that the secondary is relatively small and/or still somewhat distant. In particular, $p_{\rm ideal}$-like perturbers (near $\sim 20$ pc) can likely be ruled out as they would suppress the eccentricity gradient in the disk. Additionally, closer black holes within the disk could also be ruled out as they would disrupt its apsidal alignment. 

\section{Acknowledgements}
This work was supported by a NASA Astrophysics Theory Program under grant NNX17AK44G.  
AM gratefully acknowledges support from the David and Lucile Packard Foundation.
Simulations in this paper made use of the REBOUND code which can be downloaded freely at \hyperlink{https://github.com/hannorein/rebound}{https://github.com/hannorein/rebound}. This work utilized the RMACC Summit supercomputer, which is supported by the National Science Foundation (awards ACI-1532235 and
ACI-1532236), the University of Colorado Boulder, and Colorado State University. The Summit supercomputer is a joint effort of the University of Colorado Boulder and Colorado State University.

\textit{Software:} REBOUND~\citep{rein.liu2012}

\section*{Data Availability}
Simulation data will be made shared upon reasonable request.

\clearpage
\appendix
\section{Analytic Precession Rate}
\label{sec:Appendix}

Here we derive an approximate expression for the precession rate of a disk orbit due to an external perturber. 
To quadrupole order, the perturbing potential on the disk orbit is 

\begin{equation}
    U_p=\frac{G m_p d^2}{2 a_p^3} P_2( \cos (\theta) ),
    \label{eq:pertPot}
\end{equation}
where $M_p$ amd $a_p$ are the mass and semi-major axis of the perturber (assumed to be on a circular orbit); $d$ is the  distance between the disk particle and the central mass; $P_2$ is the second order Legendre polynomial; $\theta$ is the polar angle of the disk particle (the perturber inclination is 0). (Equation~\eqref{eq:pertPot} can be derived by averaging the quadrupole term in equation 6.23 of \citealt{murray&dermott2000} over the perturber orbit.) In terms of the Keplerian orbital elements of the disk orbit, $\cos(\theta) =\sin(\omega+f) \sin(i)$ (cf equation 2.122 in \citealt{murray&dermott2000}). Here $f$, $\omega$, $\Omega$, $e$, $a$, and $i$ are the true anomaly, argument of pericenter, longitude of the ascending node, eccentricity, semi-major axis, and inclination of the disk orbit. The average of the perturbing potential over the disk orbit is 

\begin{align}
    R &= \frac{G m_p}{2 a_p^3} \frac{ \oint d^2 P_2(\cos(\theta)) dt }{P} \nonumber\\
      &= \frac{G m_p a^4}{2 a_p^3 j} \frac{ \oint \left(\frac{d}{a}\right)^4 P_2(\cos(\theta)) d\theta }{P}\nonumber\\
      &=\frac{G m_p}{a_p} \frac{a^2}{a_p^2}\frac{\left(30 e^2 \sin ^2(i) \cos (2 w)+\left(3 e^2+2\right) (3 \cos (2 i)+1)\right)}{32}.
      \label{eq:R}
\end{align}
In the above equations $j$ and $P$ are the angular momentum and period of the disk orbit respectively. 
The total precession rate of the disk orbit is

\begin{align}
    \frac{d\varpi}{dt}&=\frac{d(\Omega\pm \omega)}{dt},
\end{align}
where the top (bottom) sign corresponds to prograde (retrograde) orbits. 
The above expression can be computed using the Lagrange planetary equations (e.g. \citealt{dosopoulou&kalogera2016}, viz.
\begin{align}
    &\frac{d \omega}{d t}= \frac{d R}{d e} \frac{(1-e^2)^{1/2}}{e n a^2}- \frac{d R}{d i} \frac{(1-e^2)^{-1/2} \cot(i)}{e n a^2}\nonumber\\
    &\frac{d \Omega}{d t}= \frac{d R}{d i} \frac{(1-e^2)^{-1/2}}{\sin(i) n a^2},
    \label{eq:lag}
\end{align}
where $n=2 \pi/P$. Combining equations~\eqref{eq:R} and~\eqref{eq:lag} gives 

\begin{align}
    \frac{d\varpi}{dt}=& \frac{3 n}{8} \left(\frac{m_p}{M}\right) \left(\frac{a}{a_p}\right)^3 \left[\frac{\cos (i) \left(5 e^2 \cos (2 \omega)-3 e^2-2\right)}{ \sqrt{1-e^2}} \mp\right. \nonumber \\
    &\left.\frac{\left(20 e^2
  \cos (2 \omega)-4 e^2+5 \cos (2 (i-\omega))+5 \cos (2 (i+\omega))-10 \cos (2 i)-10 \cos (2 \omega)-6\right)}{4 \sqrt{1-e^2}}\right]
\end{align}
For planar orbits ($i=0, \pi$) the precession rate is
\begin{align}
 \frac{d\varpi}{dt}&= \pm \frac{3}{4} \left(\frac{m_p}{M}\right) \left(\frac{a}{a_p}\right)^3 \sqrt{1-e^2} n  \nonumber\\
 &=\pm \frac{3}{2} \pi \left(\frac{m_p}{M}\right) \left(\frac{a}{a_p}\right)^3 \sqrt{1-e^2} P^{-1}
 \label{eq:prec_analytic}
\end{align}
The precession is always prograde with respect to the orbital angular momentum. Equation~\eqref{eq:prec_analytic} also corresponds to $\frac{d i_e}{d t}$ for disk stars, since $\frac{d i_e}{d t}=\frac{d \varpi}{d t}$ for planar orbits. 

\section{Numerical Precession Rate}
\label{sec:AppendixNum}
In the preceding section, we derived an analytic expression using the Lagrange planetary equations with several approximations:
\begin{enumerate}
    \item We average the perturbing potential assuming a fixed Keplerian orbit around the primary black hole.
    \item We only consider quadrupole level perturbations. 
\end{enumerate}
These approximations break down for large, close-in perturbers (though the octupole perturbation is always zero for circular perturbers).

Here, we directly measure the precession rates in REBOUND simulations. Specifically, we initialize simulations with three particles: a central black hole, a massless test particle, and a circular perturber from Table~\ref{tab:params2}. The test particles have an eccentricity of 0.7 and a semi-major axis between 1 and 2.

For each set perturber and test particle parameters, we measure the average precession rate ($\varpi'$) over fifty orbits of the test particle. To minimize the effects of two-body scatterings, we perform one hundred different simulations with randomized mean anomalies and average the results. 

\begin{figure}
    \centering
    \includegraphics[width=.5\textwidth]{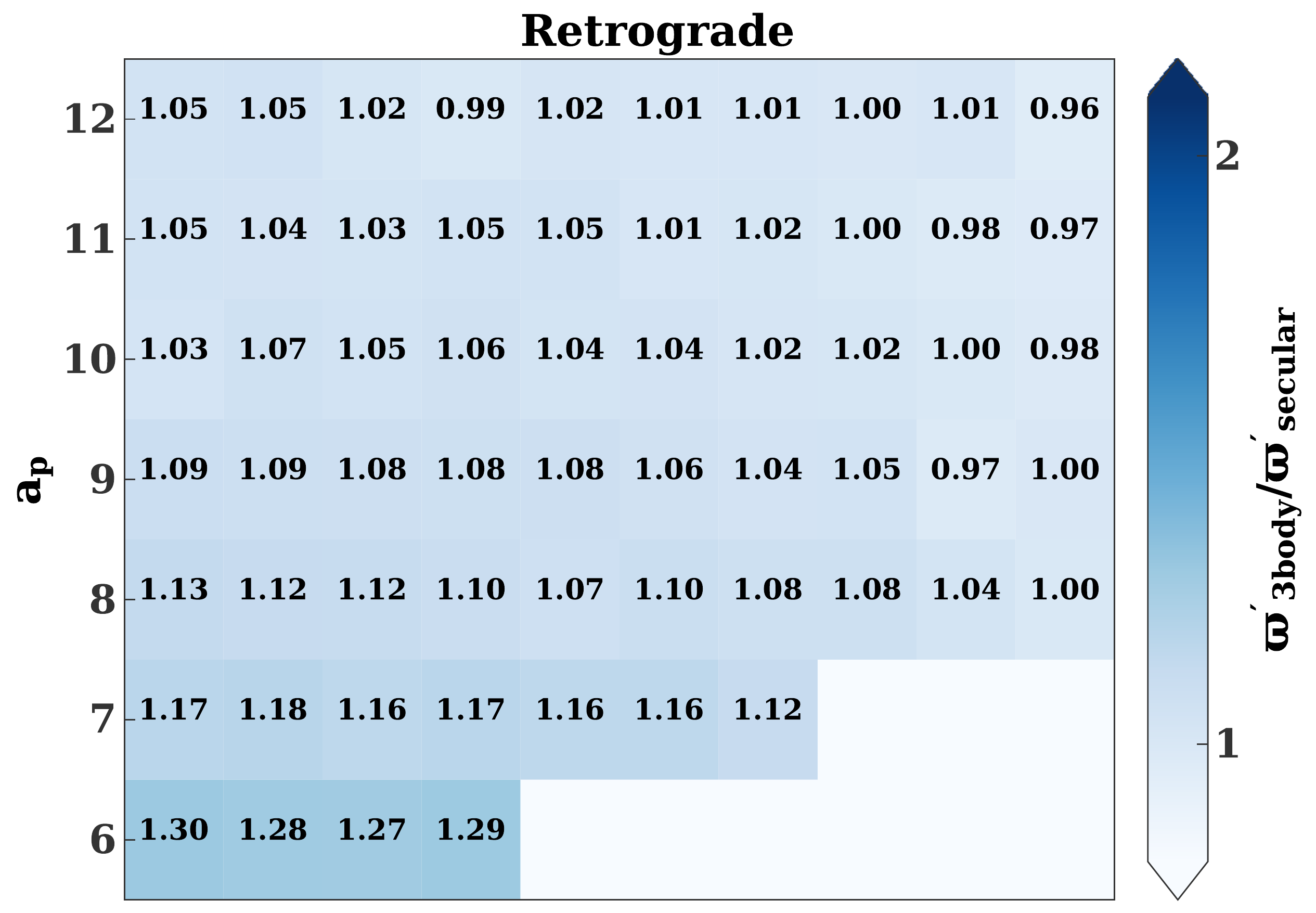}
    \includegraphics[width=.5\textwidth]{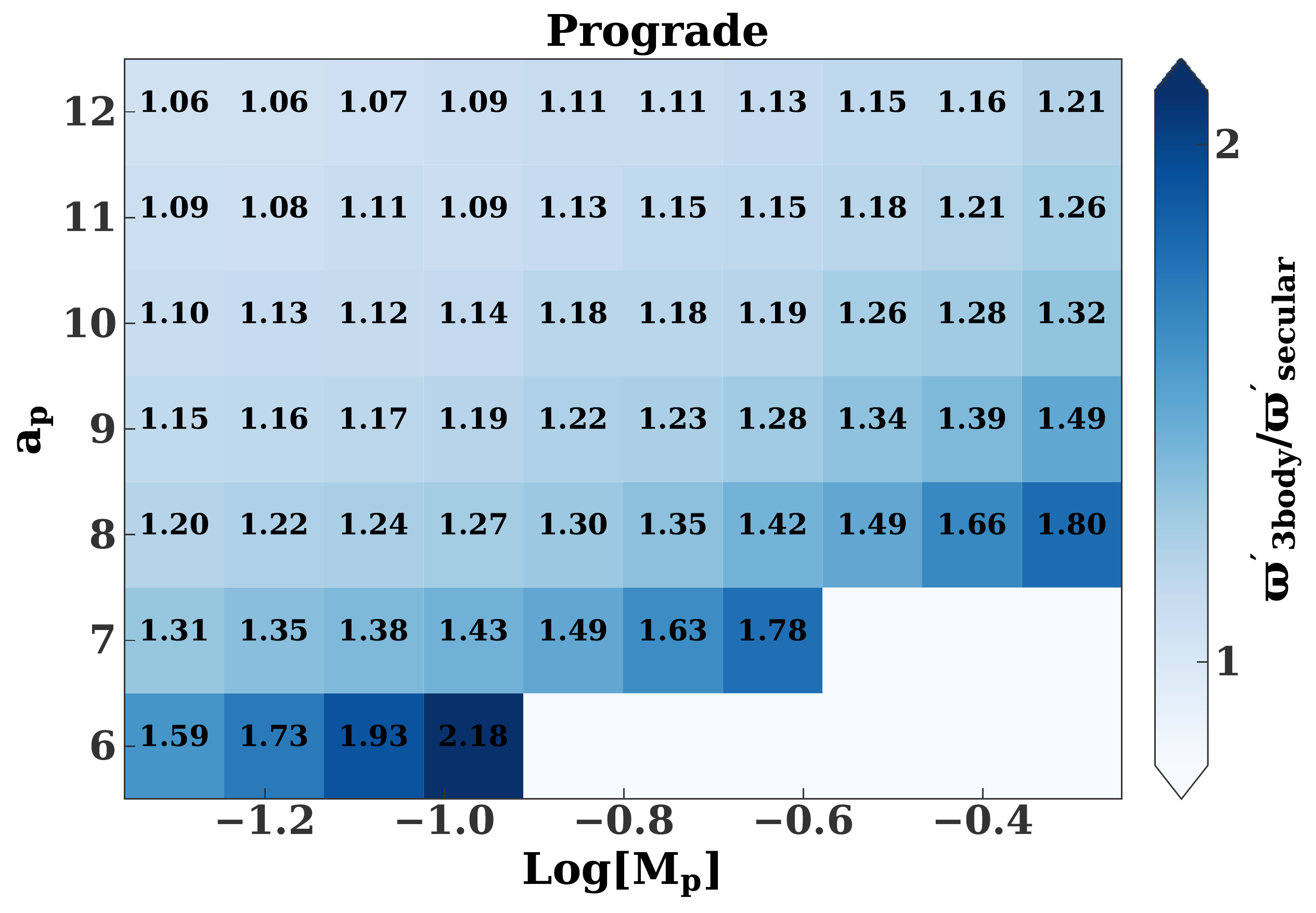}
    \caption{Precession rate of a test particle as a function perturber mass, $M_p$, and semi-major axis, $a_p$, from direct three-body integrations. The rates are normalized by equation~\eqref{eq:prec_analytic} in Appendix~\ref{sec:Appendix} (see text for details). The perturbers are coplanar with the test particle and retrograde (prograde) in the top (bottom) panel. The semi-major axis and eccentricity of the test particle are 1.9 and 0.7 respectively.}
    \label{fig:numeric_prec}
\end{figure}

Figure~\ref{fig:numeric_prec} shows the ratio of precession rates from these simulations to the analytic precession rates from Appendix~\ref{sec:Appendix} as a function of perturber parameters. For consistency with the previous appendix we use the angle $\varpi$ instead of $i_e$ to measure the precession rate here; however $d \varpi/dt$ and $d i_e/dt$ are equivalent for planar orbits. In this figure, the test particle is near the initial outer edge of the disk in our previous simulations (with a semi-major axis of 1.9). As expected, the analytic formalism is most accurate for smaller, distant perturbers, and diverges from the numerical results for small, close perturbers. In fact, for perturbers in blank regions of parameter space, the evolution of the test particle's orbit is not well described by steady precession. (For simplicity we exclude these regions from our analysis.) 

Interestingly, retrograde perturbers (top panel) are better described by the analytic formalism than prograde perturbers. This is because non-secular kicks are stronger are in the prograde case. (The relative velocity between the perturber and the test star at their closest separation is smaller for prograde perturbers than for retrograde perturbers.)

\clearpage
\footnotesize{
\bibliographystyle{mnras}
\bibliography{master}
}

% Don't change these lines
\bsp	% typesetting comment
\label{lastpage}
\end{document}